\documentclass[prd,twocolumn,superscriptaddress,nofootinbib,amsmath,%
amssymb,draft]{revtex4-1}
\usepackage[usenames]{color}
\usepackage{soul}
\usepackage{mathrsfs}

\begin{document}

\title{Exact ghost-free bigravitational waves}

\author{Eloy Ay\'on-Beato}
\email{ayon-beato@fis.cinvestav.mx}
\affiliation{Departamento de F\'isica,
CINVESTAV-IPN, Apartado Postal 14740, 07360 M\'exico D.F., M\'exico}

\author{Daniel Higuita-Borja}
\email{dhiguita@fis.cinvestav.mx}
\affiliation{Departamento de F\'isica,
CINVESTAV-IPN, Apartado Postal 14740, 07360 M\'exico D.F., M\'exico}

\author{Julio A. M\'endez-Zavaleta}
\email{jmendezz@fis.cinvestav.mx}
\affiliation{Departamento de F\'isica,
CINVESTAV-IPN, Apartado Postal 14740, 07360 M\'exico D.F., M\'exico}
\affiliation{Max-Planck-Institut f\"ur Physik (Werner-Heisenberg-Institut)\\
 F\"ohringer Ring 6, 80805 Munich, Germany}

\author{Gerardo Vel\'azquez-Rodr\'iguez}
\email{gvelazquez@astate.edu}
\affiliation{Departamento de F\'isica,
CINVESTAV-IPN, Apartado Postal 14740, 07360 M\'exico D.F., M\'exico}
\affiliation{Arkansas State University Campus Quer\'etaro, CP 76270,
Municipio Col\'on, Quer\'etaro, M\'exico}


\begin{abstract}
We study the propagation of exact gravitational waves in the ghost-free
bimetric theory. Our focus is on type-N spacetimes compatible with the
cosmological constants provided by the bigravity interaction potential, and
particularly in the single class known by allowing at least a Killing
symmetry: the AdS waves. They have the advantage of being represented by a
generalized Kerr-Schild transformation from AdS spacetime. This entails a
notorious simplification in bigravity by allowing to straightforwardly
compute any power of its interaction square root matrix, opening the door
to explore physically meaningful exact configurations. For these exact
gravitational waves the complex dynamical structure of bigravity decomposes
into elementary exact massless or massive excitations propagating on AdS.
We use a complexified formulation of the Euler-Darboux equations to provide
for the first time the general solutions to the massive version of the
Siklos equation which rules the resulting AdS-wave dynamics, using an
integral representation originally due to Poisson. Inspired by this
progress, we tackle the subtle problem of how matter couples to bigravity and,
more concretely, if this occurs through a composite metric, which is hard to
handle in a general setting. Surprisingly, the Kerr-Schild ansatz brings
again a huge simplification in how the related energy-momentum tensors are
calculated. This allows us to explicitly characterize AdS waves supported
by either a massless free scalar field or  a wavefront-homogeneous
Maxwell field. Considering the most general allowed Maxwell
source instead is a highly nontrivial task, which we accomplish by
again exploiting the complexified Euler-Darboux description and taking
advantage of the classical Riemann method. In fact, this eventually allows us to find
the most general configurations for any matter source.
\end{abstract}

\maketitle

\tableofcontents

\section{Introduction\label{Sec:Intro}}

Gravitational theories are characterized by the existence of wavy
configurations, as in any other relativistic field theory; they are responsible
for propagating the interaction and transferring energy across spacetime, which
eventually can be measured independently of where they were generated. This was
recently achieved by the LIGO and Virgo
Collaborations with the direct detection of gravitational waves for the first
time \cite{Abbott:2016blz,Abbott:2016nmj,Abbott:2017vtc,Abbott:2017oio,%
TheLIGOScientific:2017qsa,Abbott:2017gyy}, which provided the missing crucial
element in Einstein's legacy to understand the gravitational interaction as
geometry and was awarded the 2017 Nobel Prize in Physics
\cite{Nobel:2017}. However, other open questions remain to be answered, such
as the dark matter and dark energy problems, which might be tackled by
modifying General Relativity. Therefore, upcoming experimental data require
theoretical results capable of illuminating the way to proceed. In this sense,
the full understanding of gravitational-wave solutions is promising, since
they contain fingerprints to track down features of different theories,
especially beyond the perturbative treatment.

In four dimensions, one of these modified theories is the massive gravity of
de~Rham, Gabadadze, and Toley (dRGT) \cite{deRham:2010kj}. It has received a
lot of attention for being a consistent way to include nonlinear massive
gravitons in a manner that is free of the so-called Boulware-Deser ghosts
\cite{Boulware:1973my}, which permeated many previous attempts. These efforts
to include massive gravitons not only followed the theoretical insight of
Fierz and Pauli \cite{Fierz:1939ix}---through a long endeavor to consistently
incorporate self-interactions for a massive spin-2 particle\textemdash but they also rely on
the belief that such massive gravitons can help to explain some currently incomprehensible characteristics of the observed Universe.

The dRGT theory requires a reference metric in addition to the dynamical
massive one in order to formally restore diffeomorphism invariance, which
entails a precise nonderivative self-interaction potential built with the
square root matrix of the product of both metrics. As a consistent extension,
Hassan and Rosen \cite{Hassan:2011zd} showed that the reference metric can be
promoted to an independent dynamical one without spoiling the nice features
of the original theory (i.e., it is generally covariant and ghost-free), which is the why the resulting theory is called bigravity. This is the framework we
will work with, so will devote the following section to reviewing its main
ingredients.

The first gravity theory propagating a ghost-free massive degree of freedom (d.o.f.)
was Topologically Massive Gravity in three dimensions
\cite{Deser:1981wh,Deser:1982vy}. It relies in a parity-violating
Chern-Simons description, which was superseded by later proposals as New
Massive Gravity \cite{Bergshoeff:2009hq} or Zwei-Dreibein Gravity
\cite{Bergshoeff:2013xma}. The latter are theories which in the
perturbative regime recover the Fierz-Pauli three-dimensional limit; in this
sense, they are analogous to dRGT and bigravity theories in lower dimensions. All of these
three-dimensional theories support exact gravitational-wave configurations
that exhibit the particular dynamics of each theory beyond the perturbative
level, and consist of the pure gauge modes of standard $(2+1)$ gravity plus
nontrivial contributions reflecting their respective massive excitations
\cite{AyonBeato:2004fq,AyonBeato:2005bm,AyonBeato:2005qq,AyonBeato:2009yq,
Bergshoeff:2014eca} (see also the recent review \cite{Garcia:2017}).
It is expected to have similar revealing behavior in bigravity, i.e., that
exact gravitational waves decompose the complex dynamical structure of the
theory into elementary exact massless and massive excitations.

Exact gravitational waves over flat spacetime rigged by bigravity were
previously investigated in Ref.\ \cite{Mohseni:2012ug} under the restriction that
both metrics have the same profile. As we will see in the next section, the
interaction terms inherently include cosmological constants for both metrics;
hence, it is natural to search for wave solutions that can be interpreted
as being propagated over (anti--)de Sitter [(A)dS] backgrounds. One such configuration is
what is known in the literature as AdS waves, which were exhaustively studied in
General Relativity by Siklos \cite{Siklos:1985}. They can be represented by a
generalized Kerr-Schild transformation from AdS spacetime. Fortunately, it
has been proved by some of the authors that dealing with such transformations
provides a notorious simplification in bigravity, allowing to straightforwardly
compute the interaction square root matrix and any of its powers,
independently of the seed metric \cite{Ayon-Beato:2015qtt}.
Section~\ref{Sec:EGwaves} reviews the appearance of exact gravitational waves in
General Relativity and ends by motivating the AdS-wave ansatz, including its
residual symmetries and the reductions they induce in the solutions. Later,
Sec.~\ref{Sec:AdSwaves} deals with the study of AdS waves in bigravity and
how to decouple the resulting partial differential equations system for both
independent wave profiles. It turns out that one of the decoupled profiles
obey a massless Klein-Gordon equation on AdS, which is just the so-called
Siklos equation \cite{Siklos:1985}. The other satisfies a massive
Klein-Gordon equation on AdS, with an effective mass proportional to the
Fierz-Pauli one, defining a massive deformation of the Siklos equation. As a
warm up, we start in Sec.~\ref{Sec:SeparableSol} by analyzing sum-separable
solutions of these equations to get a clear decomposition in the involved
physical modes and gain intuition about the resulting configurations.
Thereafter, we address the general setting in Sec.~\ref{Sec:GenSolSik} by
regarding the Siklos operator as a complexified version of Euler-Darboux
operators \cite{Koshlyakov,Copson}, which allows us to find the most general
AdS waves rigged by the dynamics of bigravity. Another controversial issue of
bigravity is the way to couple matter, and Sec.~\ref{Sec:Matter} is devoted to
this issue with a discussion on the effective metric and explicit
calculations for both scalar and Maxwell fields. Their general solutions are
also characterized by again exploiting the Euler-Darboux description and
making use of the Riemann method, which in fact allows us to find the general
solutions of the involved exact excitations for any matter source. Final
remarks and perspectives of the present work appear in
Sec.~\ref{Sec:conclusions}. The case of \emph{pp}-waves propagating in flat
spacetime is additionally analyzed by completeness in Appendix \ref{App:pp-waves}.
Other appendices are devoted to detailed derivations which are essential to
obtain the main results of the paper.

\section{The Bigravity theory\label{Sec:Bigravity}}

Bigravity as formulated by Hassan and Rosen \cite{Hassan:2011zd} is a
four-dimensional ghost-free theory describing two dynamical metric fields
$g_{\mu\nu}$ and $f_{\mu\nu}$ interacting via a nonderivative potential,
originally proposed by dRGT to consistently describe a single massive spin-2
field \cite{deRham:2010kj}. One of the metric fields is massive and the other
is massless; hence, the theory consequently propagates a total of $5+2$
d.o.f. The interaction is encoded through terms computed from a
matrix defined by the following quadratic relation
\begin{equation}\label{eq:gamma}
{(\gamma^2)^\mu}_\nu={\gamma^\mu}_\alpha{\gamma^\alpha}_\nu
\equiv g^{\mu\alpha}f_{\alpha\nu}.
\end{equation}
The action defining bigravity is
\begin{align}
S_{\text{bi}}[g,f]={}&  \frac{1}{2\kappa_g}\int d^4x\sqrt{-g}R[g]
+\frac{1}{2\kappa_f}\int
d^4x\sqrt{-f}\mathcal{R}[f]\nonumber\\
        & -\frac{m^2}{\kappa}\int d^4x\sqrt{-g}\,\mathcal{U}[g,f],
   \label{eq:Bigravity}
\end{align}
where $m$ is the graviton mass, $R[g]$ and $\mathcal{R}[f]$ are the scalar
curvatures associated to the metrics $g_{\mu\nu}$ and $f_{\mu\nu}$,
respectively, $\kappa_{g}$ and $\kappa_{f}$ are their
corresponding gravitational constants, and $\kappa$ is in general a function
of these constants. The interaction is specified by the potential
\begin{equation}\label{eq:Int}
\mathcal{U}[g,f]=\sum_{k=0}^{4}b_k\mathcal{U}_k(\gamma),
\end{equation}
where $b_{k}$ are the coupling constants of the theory and each term of the
potential is defined as
\begin{align}
\mathcal{U}_0(\gamma)&=1,\quad
\mathcal{U}_1(\gamma) =[\gamma],\quad
\mathcal{U}_2(\gamma) =\frac{1}{2!}\!
\left([\gamma]^2-[\gamma^2]\right)\!,\nonumber\\
\mathcal{U}_3(\gamma)&=\frac{1}{3!}\!
\left([\gamma]^3-3[\gamma][\gamma^2]+2[\gamma^3]\right)\!,
\nonumber\\
\mathcal{U}_4(\gamma)&=\frac{1}{4!}\!
\left([\gamma]^4-6[\gamma]^2[\gamma^2]+8[\gamma][\gamma^3]
+3[\gamma^2]^2-6[\gamma^4]\right)\!,\label{eq:Pot}
\end{align}
where $[\gamma^{n}]$ stands for the trace of the $n$th power of the square
matrix $\gamma$ defined in Eq.~\eqref{eq:gamma}. Here, the first contribution
plays the role of a cosmological constant for the metric $g_{\mu\nu}$.
Additionally, the last one can be compactly rewritten in the action
\eqref{eq:Bigravity} as a constant times $\sqrt{-f}$, and thus it plays the role of
a cosmological constant for the other metric $f_{\mu\nu}$.

The action principle yields a set of two coupled Einstein field equations
\begin{equation}\label{eq:FieldEq}
{G^\mu}_\nu - \frac{m^2\kappa_g}{\kappa}{V^\mu}_\nu=0, \qquad
{\mathcal{G}^\mu}_\nu - \frac{m^2\kappa_f}{\kappa}{\mathcal{V}^\mu}_\nu=0,
\end{equation}
where ${G^\mu}_\nu$ and ${\mathcal{G}^\mu}_\nu$ are the Einstein tensors for
$g_{\mu\nu}$ and $f_{\mu\nu}$, respectively. The interaction tensors are
defined by varying the potential with respect to both metrics
\begin{subequations}\label{eq:interactions}
\begin{align}
{V^\mu}_\nu & \equiv \frac{2g^{\mu\alpha}}{\sqrt{-g}}\frac{\delta}{\delta
g^{\alpha\nu}}\left(\sqrt{-g}\mathcal{U}\right)=
{\tau^\mu}_\nu-\mathcal{U}{\delta^\mu}_\nu,
\label{eq:Interaction_g}\\
{\mathcal{V}^\mu}_\nu & \equiv
\frac{2f^{\mu\alpha}}{\sqrt{-f}}\frac{\delta}{\delta
f^{\alpha\nu}}\left(\sqrt{-g}\mathcal{U}\right)=
-\frac{\sqrt{-g}}{\sqrt{-f}}{\tau^\mu}_\nu,
 \label{eq:Interaction_f}
\end{align}
where
\begin{align}
{\tau^\mu}_\nu={}&
\left(b_1\mathcal{U}_0+b_2\mathcal{U}_1+b_3\mathcal{U}_2
+b_4\mathcal{U}_3\right){\gamma^\mu}_\nu
\nonumber \\
 &-\left(b_2\mathcal{U}_0+b_3\mathcal{U}_1
+b_4\mathcal{U}_2\right){\left(\gamma^2\right)^\mu}_\nu
\nonumber \\
 &+\left(b_3\mathcal{U}_0+b_4\mathcal{U}_1\right)
{\left(\gamma^3\right)^\mu}_\nu \nonumber \\
 &-b_4\mathcal{U}_0{\left(\gamma^4\right)^\mu}_\nu.
\label{eq:Interaction_tau}
\end{align}
\end{subequations}
In principle, all of the coupling constants $b_k$ are left free. But, as we
already denote the graviton mass by $m$, in the consistent linear limit
enjoyed by the theory the latter constant must correspond to the Fierz-Pauli
mass in flat space. This imposes a constraint between the
coupling constants \cite{Ayon-Beato:2015qtt}
\begin{equation}
b_2=-1-2b_3-b_4,
\label{eq:Cs2Bs}
\end{equation}
which we shall use in the present work.

\section{Exact gravitational waves: the case of AdS waves
\label{Sec:EGwaves}}

Gravitational waves are most commonly understood as small perturbations to a
background spacetime in the form $g_{\mu\nu}=g^{(0)}_{\mu\nu}+h_{\mu\nu}$,
where the quadratic and higher contributions of $h_{\mu\nu}$ are neglected,
which linearizes the intrinsically nonlinear Einstein
field equations and forces the perturbations to satisfy the standard wave
equation. This is the kind of gravitational waves recently detected by the
LIGO and Virgo observatories
\cite{Abbott:2016blz,Abbott:2016nmj,Abbott:2017vtc,Abbott:2017oio,%
TheLIGOScientific:2017qsa,Abbott:2017gyy}. Remarkably, in spite of its
inherent nonlinearity, General Relativity also allows the existence of
\emph{exact} gravitational waves. These are solutions of the Einstein equations
for which the linearization leading to the wave equation does not rely on any
approximation, but rather emerges from a different mechanism usually involving the
existence of a principal null direction \cite{Stephani:2003}. When such a
vector field associated to the Weyl tensor presents multiplicity, this defines
the so-called algebraically special spacetimes. In the case where this
multiplicity is maximal (fourfold), the spacetimes are classified as type $N$ and characterize the exact gravitational waves. If these null vector fields are
additionally geodesic (optical rays),\footnote{It is not necessary to assume
this and other properties of the null congruences under certain conditions
summarized in the celebrated Goldberg-Sachs theorem \cite{Goldberg:1962} and
its generalizations \cite{Stephani:2003,Krasinski:2009}.} they are classified
in terms of the irreducible contributions to the projection of its covariant
derivative on the two-dimensional spatial sections orthogonal to them. When
the antisymmetric part of this projection vanishes (nontwisting rays) and
the traceless contribution of the symmetrical part is also zero (shear-free
rays), there are only two possibilities: the trace also vanishes
(nonexpanding rays), or it is nontrivial (expanding rays). In vacuum, both
cases are described by the Kundt \cite{Kundt:1961} and Robinson-Trautman
\cite{Robinson:1960zzb} classes of exact gravitational waves, respectively.
The nonexpanding Kundt class contains a subcase where the multiple principal
null direction is additionally a Killing field and therefore a covariantly
constant vector, which describes \emph{plane}-fronted gravitational waves
with \emph{parallel} rays, or in short \emph{pp}-waves. This was one of the
early exact examples of gravitational waves \cite{Brinkmann:1925fr} and
probably the most studied. Their dynamics under bigravity is explored in
Appendix~\ref{App:pp-waves}.

As emphasized in the previous section, the theory we focus on in this
paper naturally presents a pair of cosmological constants. Hence, it is more
appropriate to study the kind of exact gravitational waves that can be
propagated under these circumstances. In General Relativity, the generalization
of the Kundt and Robinson-Trautman waves in the presence of a cosmological
constant was realized by the CINVESTAV group in the early 1980's
\cite{Garcia:1981,Salazar:1983,Garcia:1983}, (see also Ref.~\cite{Ozsvath:1985qn}
and the review \cite{Bicak:1999h}). Here again, the generalized Kundt class
has a subcase where the non-expanding ray becomes a Killing vector and which
was exhaustively studied by Siklos \cite{Siklos:1985}. The symmetrical Siklos
spacetimes only exist for a negative cosmological constant and are defined by
the metric
\begin{equation}\label{eq:Siklos}
ds^2=\frac{\ell^{2}}{y^{2}}\left[-F(u,y,x)du^{2}-2dudv+dy^{2}+dx^{2}\right],
\end{equation}
where the null Killing field is $\partial_v$. In General Relativity the
gravitational profile $F$ satisfies the wave equation on AdS
space written in Poincar\'e coordinates
\begin{equation}\label{eq:AdS}
ds_\text{AdS}^2=\frac{\ell^{2}}{y^{2}}\left(-2dudv+dy^{2}+dx^{2}\right),
\end{equation}
where $u$ and $v$ play the role of retarded and advanced times, respectively.
This permits the interpretation that these solutions are exact
gravitational waves propagating on an AdS background \cite{Podolsky:1997ik} or,
in short AdS waves \cite{AyonBeato:2005qq}.

The linearization that gives rise to the wave equation arises in this case because the
metric can be written as a generalized Kerr-Schild transformation
\begin{equation}\label{eq:KerrSchild}
ds^2=ds_\text{AdS}^2-F \, k \otimes k,
\end{equation}
where the vector field
\begin{equation}\label{eq:k}
k_{\mu}dx^{\mu}=-\frac{\ell}{y}du,
\end{equation}
is proportional to the Killing vector $\partial_v$ and retains its null and
geodesic properties on AdS. Satisfying such properties on any seed metric
assures that the mixed components of the transformed Ricci tensor depend
linearly on the profile $F$ and their derivatives \cite{Stephani:2003}. This
explains the emergence of the wave operator.

The line element \eqref{eq:Siklos} is form invariant under the family of
transformations
\begin{subequations}\label{eq:ressym}
\begin{align}
\tilde{u}={}&\int{\frac{du}{\mathsf{f}^{2}}},\qquad
\tilde{y}=\frac{\lambda}{\mathsf{f}}y,\qquad
\tilde{x}=\frac{\lambda}{\mathsf{f}}(x+P),\nonumber\\
\tilde{v}={}&\lambda^2\Biggl\{
          v-\frac{1}{2}\frac{\dot{\mathsf{f}}}{\mathsf{f}}(y^{2}+x^{2})
+\mathsf{f}\frac{d}{du}
\left(\frac{P}{\mathsf{f}}\right)x\nonumber\\
&\qquad+\frac12\int du\bigg[B_{0}+\dot{P}^{2}-\dot{\mathsf{f}}\frac{d}{du}
\left(\frac{P^2}{\mathsf{f}}\right)\bigg]\Biggr\},\nonumber\\
\tilde{F}={}&(\lambda\mathsf{f})^2
          \left[F-B_{2}(y^{2}+x^{2})-B_{1}x-B_{0}\right],
\label{eq:ressym1}
\end{align}
where $\mathsf{f}=\mathsf{f}(u)$, $P=P(u)$, and $B_{0}=B_{0}(u)$ are arbitrary
functions of the retarded time, a dot denotes a derivative with respect to
$u$ and the coefficients of the quadratic and linear terms in the wavefront
coordinates at the profile transformation are determined from the above
functions by
\begin{equation}\label{eq:ressym2}
B_{2}=-\frac{\ddot{\mathsf{f}}}{\mathsf{f}}, \qquad
B_{1}=2(\ddot{P}+P B_{2}).
\end{equation}
\end{subequations}
These transformations determine the \emph{residual symmetries} of the AdS
waves (\ref{eq:Siklos}), and they were discussed originally by Siklos in Ref.
\cite{Siklos:1985} and were extended to any dimension in Ref.
\cite{Ayon-Beato:2015xsz}. They can be exploited in the following way (see
Ref.~\cite{AyonBeato:2005qq}): if the solution profile contains a quadratic
term in $x$ and $y$ with the same coefficient, a linear term in $x$, and/or a
zero-order term, one can choose the functions in the transformation in order
that $B_2$, $B_1$, and/or $B_0$ coincide with the coefficients of these terms,
respectively. These selections entail differential equations for $\mathsf{f}$
and $P$ through Eq.~(\ref{eq:ressym2}), whose solutions define precise local
coordinate transformations that eliminate the involved contributions in the
transformed profile. This feature will be of great help, as we will see in
Sec.~\ref{Sec:SeparableSol} and later.

\section{A\MakeLowercase{d}S waves in Bigravity\label{Sec:AdSwaves}}

In this work we undertake the task of studying the dynamics of AdS waves in
bigravity. Hence, we will take both metrics as a generalized Kerr-Schild ansatz
of the form (\ref{eq:KerrSchild}), additionally allowing the presence of a
global conformal factor in the second metric
\begin{align}\label{eq:AdSansatz}
g_{\mu\nu}&=g_{\mu\nu}^\text{AdS}-F_{1}(u,x,y)k_{\mu}k_{\nu},\nonumber \\
f_{\mu\nu}&=C^{2}\left[g_{\mu\nu}^\text{AdS}-F_{2}(u,x,y)k_{\mu}k_{\nu}\right],
\end{align}
which provides the freedom to use two different AdS radii, whose ratio is
precisely $C=\ell_{f}/\ell_{g}$.

In the context of bigravity, not only   does the Kerr-Schild ansatz provide the
well-known linearization leading to the exact wavy behavior of General
Relativity described in the previous section, but (as was noticed first in Ref.~\cite{Ayon-Beato:2015qtt} by some of the authors) the null character of the
vector \eqref{eq:k} in the generalized transformation also provides a
nilpotent contribution to the interaction square root matrix \eqref{eq:gamma}\footnote{For preliminary results where both metrics are related by a Kerr-Schild ansatz see Ref.~\cite{Baccetti:2012ge}.}. Thus we can immediately write down the latter for any seed metric as
\begin{equation}
{\gamma^\mu}_\nu=C\left[{\delta^\mu}_\nu
-\frac{1}{2}(F_{2}-F_{1})k^{\mu}k_{\nu}\right],
\label{eq:gammaKerrSchild}
\end{equation}
where $k^\mu\equiv g^{\mu\nu}k_\nu$. This property allows to trivially
calculate the powers of the interaction matrix necessary to compute the entire set of
field equations
\begin{equation}
{(\gamma^n)^\mu}_\nu=C^n\left[{\delta^\mu}_\nu
-\frac{n}{2}(F_{2}-F_{1})k^{\mu}k_{\nu}\right].
\label{eq:gamma2n}
\end{equation}
A long but straightforward calculation shows that the interaction tensors
acquire an almost diagonal form, except for a contribution along the null ray
\eqref{eq:k}, meaning that the field equations (\ref{eq:FieldEq}) take the
simple form
\begin{subequations}\label{eq:EinsteinEqs}
\begin{align}
{G^{\mu}}_{\nu}-\frac{\kappa_{g}m^2}{\kappa}\left(
P_{1}{\delta^\mu}_\nu-C\frac{P_0}{2}(F_{2}-F_{1})k^{\mu}k_{\nu} \right)&=0,
\label{eq:Eg}\\
{\mathcal{G}^{\mu}}_{\nu}-\frac{\kappa_{f}m^2}{\kappa C^3}\left(
P_{2}{\delta^\mu}_\nu+\frac{P_{0}}{2}(F_{2}-F_{1})k^{\mu}k_\nu \right)&=0,
\label{eq:Ef}
\end{align}
where the dependence on the coupling constants is encoded in the following
combinations
\begin{align}
P_0&\equiv -2Cb_4+C(C-4)b_3+b_1-2C,\label{eq:P0}\\
P_1&\equiv 3C^2b_4-C^2(C-6)b_3-3Cb_1-b_0+3C^2,\label{eq:P1}\\
P_2&\equiv -C(C^2-3)b_4-3C(C-2)b_3-b_1+3C.\label{eq:P2}
\end{align}
\end{subequations}

Each Einstein tensor only contributes to the diagonal with a term
proportional to the inverse of the corresponding AdS radius
\begin{subequations}\label{eq:EinsteinAdSWaves}
\begin{align}
{G^{\mu}}_{\nu}-3\ell^{-2}{\delta^{\mu}}_{\nu}\propto k^{\mu}k_{\nu},\\
{\mathcal{G}^{\mu}}_{\nu}-3C^{-2}\ell^{-2}{\delta^{\mu}}_{\nu}
\propto k^{\mu}k_{\nu}.
\end{align}
\end{subequations}
Hence, the two diagonal contributions from the Einstein equations
\eqref{eq:EinsteinEqs} fix two combinations of coupling constants in terms of
the AdS radii, defining an effective negative cosmological constant for each
metric
\begin{subequations}\label{eq:AdSRelation}
\begin{align}
\Lambda^\text{eff}_1&\equiv-\frac{\kappa_{g}m^2}{\kappa}P_{1}
=-\frac{3}{\ell^{2}}, \\
\Lambda^\text{eff}_2&\equiv-\frac{\kappa_{f}m^2}{\kappa C^3}P_{2}
=-\frac{3}{C^2\ell^{2}}.
\end{align}
\end{subequations}
The other nontrivial terms are the off-diagonal components along the null
ray, which can be written covariantly as
\begin{subequations}\label{eq:F1F2sys}
\begin{align}
\left(\frac12\frac{y^2}{\ell^2}\Delta_\text{S}F_{1}
+\frac{C\kappa_{g}m^2P_{0}}{2\kappa}(F_{2}-F_{1})\right)
k_\mu k_\nu &=0,
\label{eq:deF1}\\
\left(\frac12\frac{y^2}{\ell^2}\Delta_\text{S}F_{2}
-\frac{\kappa_{f}m^2P_{0}}{2C\kappa}(F_{2}-F_{1})\right)
k_\mu k_\nu &=0,
\label{eq:deF2}
\end{align}
\end{subequations}
where $\Delta_\text{S}$ is the operator defining the Siklos equation, defined
below in Eq.~\eqref{eq:FDE}, given by
\begin{equation}\label{def:SiklosOp}
\Delta_\text{S}\equiv\Delta_\text{L}-\frac{2}{y}\partial_{y},\qquad
\Delta_\text{L}\equiv\partial_{y}^{2}+\partial_{x}^{2},
\end{equation}
with $\Delta_\text{L}$ being the standard Laplacian operator. With the help of
the profile redefinitions\footnote{This decoupling is similar to that found
perturbatively in the cosmological context \cite{Max:2017flc}.}
\begin{subequations} \label{def:SolvableCom}
\begin{align}
\mathscr{F}&\equiv\frac{\kappa_f}{\kappa}F_1
            +\frac{C^2\kappa_g}{\kappa}F_2,\\
\mathscr{H}&\equiv F_2-F_1,
\end{align}
\end{subequations}
the system (\ref{eq:F1F2sys}) is decoupled and becomes
\begin{subequations}\label{eq:FHsys}
\begin{align}
\Delta_\text{S}\mathscr{F}&=0,\label{eq:FDE}\\
\Delta_\text{S}\mathscr{H}-\frac{\ell^{2}\hat{m}^{2}}{y^{2}}
\mathscr{H}&=0.\label{eq:HDE}
\end{align}
The first equation for $\mathscr{F}$ corresponds to the Siklos equation, since
he was the first to describe the dynamics of AdS waves \cite{Siklos:1985}. Up
to a factor, this is just the wave operator (d'Alembertian) evaluated on the AdS metric \eqref{eq:AdS}
written in Poincar\'e coordinates. In other words, the
profile $\mathscr{F}$ describes an exact massless excitation. The second
equation for $\mathscr{H}$ is the massive version of the Siklos one since
Eq.~(\ref{eq:HDE}), up to a factor, is just the massive Klein-Gordon equation
evaluated on the AdS spacetime \eqref{eq:AdS} with a mass given in terms of the Fierz-Pauli
one as follows
\begin{equation}\label{def:meff}
\hat{m}^{2}\equiv\frac{\left(\kappa_{f}+C^{2}\kappa_{g}\right)P_{0}}
{C\kappa}m^{2}.
\end{equation}
\end{subequations}
Correspondingly, the profile $\mathscr{H}$ characterizes an exact massive
excitation.

Finally, it is important to know how the exact decoupled excitations are
defined modulo diffeomorphisms in order to properly identify their physically
relevant contributions. Since we are using the same coordinates to write both
metrics ($f$ and $g$), it is easy to check that the new decoupled profiles
change under the residual symmetries \eqref{eq:ressym} of AdS waves in the
following way
\begin{subequations}\label{eq:ResSymmetryFH}
\begin{align}
\tilde{\mathscr{F}}&=(\lambda\mathsf{f})^2\!
\left\{\!\mathscr{F}\!-(\kappa_{f}+C^{2}\kappa_{g})
[B_{2}(x^2+y^2)+\!B_{1}x+\!B_{0}]\right\}\!,\label{eq:ResSymmetryF}\\
\tilde{\mathscr{H}}&=(\lambda\mathsf{f})^2\mathscr{H}.\label{eq:ResSymmetryH}
\end{align}
\end{subequations}
This means that only the massless profile inherits the characteristic
indeterminacy of the AdS waves and the massive one remains essentially the
same modulo a trivial scaling.

In the following we shall characterize these decoupled exact excitations,
first by inspecting their sum-separable sector in order to identify the
principal modes ruling the dynamics, and later by unveiling their full space of
solutions using Euler-Darboux operators. From the obtained behaviors the
original AdS-wave profiles can be reconstructed by inverting the redefinitions
\eqref{def:SolvableCom} as
\begin{subequations}\label{def:F1F2}
\begin{align}
F_{1}&=\frac1{\kappa_f+C^2\kappa_g}
\left(\kappa\mathscr{F}-C^2\kappa_g\mathscr{H}\right),\\
F_{2}&=\frac1{\kappa_f+C^2\kappa_g}
\left(\kappa\mathscr{F}+\kappa_f\mathscr{H}\right).
\end{align}
\end{subequations}

\section{Separable configurations \label{Sec:SeparableSol}}

It is very illustrative to start studying the wavefront-coordinates
sum-separable solutions to the decoupled exact excitations (\ref{eq:FHsys}).
This procedure provides a quick integration and, more importantly, reveals the
d.o.f. propagated by the theory remaining on these configurations
which helps to gain intuition on the exhibited physical modes as well as the
special points of the parameter space of the theory.

\subsection{Massless profiles $\hat{m}=0$ \label{Subsec:massless_sep}}

The closest  case to General Relativity corresponds to the vanishing of the
effective mass \eqref{def:meff}, $\hat{m}=0$. A particularly
interesting possibility is having zero-mass modes without requiring the
flat-space graviton mass $m$ to vanish, but rather imposing $P_{0}=0$. This
corresponds to a constraint on the coupling constants of the theory that
reduces the coupled differential system (\ref{eq:F1F2sys}) to the pair of
decoupled Siklos equations
\begin{equation}\label{eq:maslessFHsys}
\Delta_\text{S}F_{1}=0=\Delta_\text{S}F_{2}.
\end{equation}
As explained before, by looking for a clear decomposition in the prevailing modes of
the theory we will search for solutions that are sum separable with respect to the
wavefront coordinates, i.e., $F_1(u,y,x)=X_{1}(u,x)+Y_{1}(u,y)$ and
$F_2(u,y,x)=X_{2}(u,x)+Y_{2}(u,y)$. As a consequence of this process, each
Siklos equation is separated into ordinary Euler equations for the wavefront
coordinates. The linearly independent solutions are power laws and together
lead to the more general separable solutions in the massless case
\begin{subequations}\label{eq:solsFHmassless}
\begin{align}
F_{1}(u,x,y)={}&f_{3}(u)\left(\frac{y}{\ell}\right)^{3}
+\frac{f_{2}(u)}{\ell^{2}}\left(x^{2}+y^{2}\right)\nonumber\\
&+f_{1}(u)\frac{x}{\ell}+f_{0}(u),\\
F_{2}(u,x,y)={}&h_{3}(u)\left(\frac{y}{\ell}\right)^{3}
+\frac{h_{2}(u)}{\ell^{2}}\left(x^{2}+y^{2}\right)\nonumber\\
&+h_{1}(u)\frac{x}{\ell}+h_{0}(u),
\end{align}
\end{subequations}
where the $f$'s and $h$'s are arbitrary functions of the retarded time $u$.
We may be tempted to use the residual symmetries \eqref{eq:ressym} to remove
all the terms that have an unphysical meaning in standard gravity
\cite{AyonBeato:2005qq}; interestingly, this can be achieved for only one of the metrics, while in general the other profile will keep all of terms
\begin{subequations}\label{eq:solsFHmassless2}
\begin{align}
F_{1}(u,x,y)={}&f_{3}(u)\left(\frac{y}{\ell}\right)^{3},\\
F_{2}(u,x,y)={}&h_{3}(u)\left(\frac{y}{\ell}\right)^{3}
+\frac{h_{2}(u)}{\ell^{2}}\left(x^{2}+y^{2}\right)\nonumber\\
&+h_{1}(u)\frac{x}{\ell}+h_{0}(u).
\end{align}
\end{subequations}
We note that $F_1$ corresponds to the physical General Relativity mode, which
in this context describes the well-known Kaigorodov spacetime
\cite{Kaigorodov}. Instead, $F_2$ includes additional contributions that cannot be dropped  and are fingerprints of the massive d.o.f. of the
present theory.

\subsection{Massive profiles $\hat{m}\neq0$\label{Subsec:massive_sep}}

Let us return to the most general case in which the effective mass is not
trivial and the AdS waves are described by the decoupled exact excitations
\eqref{def:SolvableCom} obeying the system \eqref{eq:FHsys}. Once again, in
order to exhibit the decomposition in the prevailing modes we will search for
decoupled solutions that are sum separable with respect to the wavefront
coordinates, i.e., $\mathscr{F}(u,y,x)=X_{1}(u,x)+Y_{1}(u,y)$ and
$\mathscr{H}(u,y,x)=X_{2}(u,x)+Y_{2}(u,y)$. The separation in terms of ordinary Euler
equations for the wavefront coordinates now results in the solution
\begin{subequations}\label{eq:solsFH}
\begin{align}
\mathscr{F}(u,x,y)&=f_{3}(u)\left(\frac{y}{\ell}\right)^{3},
\label{eq:solFfull}\\
\mathscr{H}(u,x,y)&=h_{+}(u)\left(\frac{y}{\ell}\right)^{\rho_{+}}
+h_{-}(u)\left(\frac{y}{\ell}\right)^{\rho_{-}}\label{eq:solHfull},
\end{align}
where
\begin{equation}\label{eq:rho+-}
\rho_\pm\equiv\frac{3}{2}\pm\ell\sqrt{\hat{m}^{2}-m_{\text{BF}}^{2}},
\end{equation}
are the roots of the characteristic polynomial determining the linearly
independent power-law solutions of the ordinary Euler equation for
the massive modes (since their separable $x$ dependence must be trivial), and
\begin{equation}\label{eq:BF}
m_\text{BF}^2\equiv-\frac9{4\ell^2},
\end{equation}
\end{subequations}
corresponds to the well-known Breitenlohner-Freedman bound, which is the
lowest value the square of the mass of a stable scalar field can take on an
AdS background \cite{Breitenlohner:1982jf}. Since the profile $\mathscr{F}$
remains massless we already exploited the residual symmetries
(\ref{eq:ResSymmetryFH}) to get rid of the unphysical terms (which are like
those of the previous subsection) and reduce its solution to Eq.~\eqref{eq:solFfull}. Hence, the profiles of both metrics that are free from irrelevant
contributions are constructed using the inversions \eqref{def:F1F2} and are given
by
\begin{subequations}\label{eq:profiles}
\begin{align}
F_1(u,x,y)={}&\frac1{\kappa_{f}+C^{2}\kappa_{g}}
\Biggl\{\kappa f_{3}(u)\left(\frac{y}{\ell}\right)^3\nonumber\\
&-C^2\kappa_{g}\Biggl[
h_{+}(u)\left(\frac{y}{\ell}\right)^{\rho_{+}}
+h_{-}(u)\left(\frac{y}{\ell}\right)^{\rho_{-}}\Biggr]\Biggr\},\\
F_2(u,x,y)={}&\frac1{\kappa_{f}+C^{2}\kappa_{g}}
\Biggl\{\kappa f_{3}(u)\left(\frac{y}{\ell}\right)^3\nonumber\\
&+\kappa_{f}\Biggl[
h_{+}(u)\left(\frac{y}{\ell}\right)^{\rho_{+}}
+h_{-}(u)\left(\frac{y}{\ell}\right)^{\rho_{-}}\Biggr]\Biggr\}.
\end{align}
\end{subequations}

The saturation of the Breitenlohner-Freedman bound,
$\hat{m}^2=m_{\text{BF}}^{2}$, leads us to a logarithmic profile due to the
multiplicity in the powers of the solutions to the involved ordinary Euler
operator, $\rho_{+}=\rho_{-}=3/2$,
\begin{subequations}
\begin{align}
\mathscr{F}(u,x,y)&=f_{3}(u)\left(\frac{y}{\ell}\right)^{3},\\
\mathscr{H}(u,x,y)&=\left(\frac{y}{\ell}\right)^{\frac{3}{2}}
\left(h_{1}(u)+h_{2}(u)\ln{\frac{y}{\ell}}\right),
\end{align}
\end{subequations}
where we have used the residual symmetry (\ref{eq:ResSymmetryFH}) to get rid
of the unphysical terms. Again, the profile functions can be explicitly
written through Eqs.~(\ref{def:F1F2}) as
\begin{subequations}
\begin{align}
F_{1}(u,x,y)={}&\frac1{\kappa_{f}+C^{2}\kappa_{g}}
\biggl[\kappa f_{3}(u)\left(\frac{y}{\ell}\right)^3\nonumber\\
&-C^2\kappa_{g}\left(\frac{y}{\ell}\right)^{\frac32}
\left(h_{1}(u)+h_{2}(u)\ln{\frac{y}{\ell}}\right) \biggr], \\
F_{2}(u,x,y)={}&\frac1{\kappa_{f}+C^{2}\kappa_{g}}
\biggl[\kappa f_{3}(u)\left(\frac{y}{\ell}\right)^3\nonumber\\
&+\kappa_{f}\left(\frac{y}{\ell}\right)^{\frac32}
\left(h_{1}(u)+h_{2}(u)\ln{\frac{y}{\ell}}\right) \biggr].
\end{align}
\end{subequations}
As we saw before, one advantage of looking for sum-separable solutions is
that it allows to implement the whole residual symmetry [Eq.~\eqref{eq:ressym} or
\eqref{eq:ResSymmetryFH}] to get rid of the nonphysical terms. This clearly
exhibits a mode coming from General Relativity as well as two additional physical massive
modes that are characteristic of these kind of theories; these are the prevailing
modes respecting the symmetries of the system. For completeness, the same
approach is applied to the pp-wave problem in Appendix~\ref{App:pp-waves},
providing new results which extend those already reported for massive gravity
\cite{Mohseni:2012ug}.

\section{(Massive) Siklos excitations from Euler-Darboux equations
\label{Sec:GenSolSik}}

\subsection{General exact massless excitations}

In order to find the solution to the system (\ref{eq:FHsys}) in a general setting,
it is quite convenient to incorporate the wavefront coordinates in a complex
variable $z=x+iy$ together with its complex conjugate $\bar{z}$, in terms of
which the Siklos operator \eqref{def:SiklosOp} can be cast into the form
\begin{equation}\label{eq:ComplexFDE}
\frac{1}{4}\Delta_\text{S}=\partial^{2}_{z\bar{z}}
-\frac{1}{z-\bar{z}}\partial_{\bar{z}}
+\frac{1}{z-\bar{z}}\partial_{z}.
\end{equation}
The right-hand side is just a complexified version of the Euler-Darboux
operator $E_{\alpha,\beta}$, which we review in Appendix~\ref{App:ED} (see also
Refs.~\cite{Koshlyakov,Copson}), in the case where the parameters of the operator \eqref{eq:ede} take the values $\alpha=-1=\beta$. The complex variables imply that
the Euler-Darboux operator $E_{\alpha,\beta}$ is no longer hyperbolic, but rather
elliptic. In other words, the Siklos equation (\ref{eq:FDE}) is just a
complexified subclass of the Euler-Darboux differential equations
\cite{Siklos:1985}.\footnote{More precisely, when the two parameters $\alpha$
and $\beta$ coincide it is called an Euler-Poisson-Darboux equation.}
Consequently, the general solution of the Siklos equation can be
straightforwardly read from the related expression (\ref{eq:solgen2}) when the parameters of the
Euler-Darboux operator are negative integers and
using the reality condition on the solution. Concretely, for the Siklos case,
$\alpha=-1=\beta$ we end with
\begin{align}\label{eq:solED1}
\mathscr{F}(u,z,\bar{z}) & = \mathfrak{u}(-1,-1)\nonumber\\
& =\left(z-\bar{z}\right)^3\frac{\partial^2}{\partial{z}\partial{\bar{z}}}
\left(\frac12\frac{\omega(u,z)+\overline{\omega(u,z)}}{z-\bar{z}}\right)
\nonumber\\
& = y^2\partial_{y}\!\left(\frac{\omega(u,z)+\overline{\omega(u,z)}}{y}
\right)\!,
\end{align}
for an arbitrary complex function $\omega(u,z)$, holomorphic on the complex
wavefront coordinate $z$. This is just the general solution first reported by
Siklos in Ref.~\cite{Siklos:1985}. The connection between the Siklos equation and
the Euler-Darboux operators was pointed out by Siklos himself
\cite{Siklos:1985}. But he did not exploit this fact since he arrived at the solution
 in a different and clever way by using the following third-order
identity, satisfied for any function $f$, that allows another representation
of the Siklos operator in terms of the Laplacian
\begin{equation}\label{eq:SiklosIdentity}
\Delta_\text{S}\!\left(y^{2}\partial_{y}f \right)=
y^2\partial_{y}\!\left(\frac{1}{y}\Delta_\text{L}\!\left(yf \right) \right)\!.
\end{equation}
This allows to use the well-known fact that the real part of an arbitrary
holomorphic function is harmonic and the general solution to the Laplace equation
can always be represented in this way. In this sense, it is not strictly necessary
to invoke the alternative derivation we present first. However, as we shall
see in the following subsection, the former view is essential to study the
general solutions of the massive case for which no previous results are
known.

Before that, we need to reanalyze the manifestation of residual symmetries
since only the massless modes are practically sensitive to them, see Eq.\ \eqref{eq:ResSymmetryFH}. Such symmetries are encoded in
the holomorphic function $\omega(u,z)$, implying that it must necessarily
change under the transformation \eqref{eq:ressym} according to
\begin{equation}\label{eq:ressym_omega}
\tilde{\omega}=(\lambda\mathsf{f})^2\left(\omega+\frac{\kappa_{f}
+C^{2}\kappa_{g}}2(B_2z^{2}+B_1z+B_0)\right).
\end{equation}
Hence, any quadratic, linear, and constant holomorphic dependences on the
wavefront coordinates having real coefficients depending on retarded time $u$
can be eliminated from $\omega(u,z)$ with the help of residual symmetries.

\subsection{General exact massive excitations}

The equation defining the exact massive modes (\ref{eq:HDE}) can be written
in complex wavefront variables as
\begin{equation}\label{eq:ComplexHDE}
\left(\partial^{2}_{z\bar{z}}
-\frac{1}{z-\bar{z}}\partial_{\bar{z}}
+\frac{1}{z-\bar{z}}\partial_{z}
+\frac{\ell^2\hat{m}^2}{(z-\bar{z})^2}\right)\mathscr{H}=0,
\end{equation}
where the effective mass is given by Eq.~\eqref{def:meff}. This massive
generalization is no longer described by an Euler-Darboux equation, but
fortunately (as we review at the end of Appendix~\ref{App:ED}) it is connected to
the extension \eqref{eq:extendedEDE} of Euler-Darboux operators containing the
original massless version and whose behavior can be determined again in terms
of the standard Euler-Darboux description. This is achieved by means of the
redefinition
\begin{subequations}\label{eq:H2h}
\begin{equation}\label{eq:H2hwr}
\mathscr{H}(u,z,\bar{z})=\left(\frac{z-\bar{z}}{2i}\right)^{\rho}h(u,z,\bar{z}),
\end{equation}
where
\begin{equation}\label{eq:rho}
\rho=\rho_\pm=\frac{3}{2}\pm\ell\sqrt{\hat{m}^{2}-m_{\text{BF}}^{2}},
\end{equation}
\end{subequations}
is any of the roots defined in \eqref{eq:rho+-}, which remarkably allows us to
rewrite the massive equation as
\begin{equation}\label{eq:massiveredef}
\left(\partial^{2}_{z\bar{z}}
+\frac{\rho-1}{z-\bar{z}}\partial_{\bar{z}}
-\frac{\rho-1}{z-\bar{z}}\partial_z\right)h=0.
\end{equation}
As a consequence, the nonderivative massive contribution is dropped at the
cost of changing the kinetic parameters to $\alpha=\beta=\rho-1$, which of
course for a generic value of mass gives nothing more than the general form
of an Euler-Poisson-Darboux equation. Notice the special selection of the
constant coefficient in front of the right-hand side of redefinition
\eqref{eq:H2hwr}; in contrast to the treatment in Appendix~\ref{App:ED} where all the variables in
Eq.~\eqref{eq:U2u} are real, here we need to guarantee that the final result is real. It is necessary to remark that (as is proven in Appendix~\ref{App:ED}) we can
equivalently represent the massive configuration in terms of one root or the
other, which is why we refer to them generically as $\rho$ in Eq. \eqref{eq:rho}.
In Appendix~\ref{App:ED} we address how to find the general solution to these
equations in terms of a superposition constructed by means of an integral
representation. However, before addressing the general problem with the
methods described in Appendix~\ref{App:ED}, we find it illustrative to first study a
particular class for which the profile can be written without the use of
integrals; this case involves a discretization of the mass above the
Breitenlohner-Freedman bound \eqref{eq:BF}.

\subsubsection{Special case with discrete mass}

We start by assuming that the parameters of the Euler-Poisson-Darboux
equation \eqref{eq:massiveredef} take integer values, that is, $\rho-1=n$
with $n\in\mathbb{Z}$; inserting this into the definition of the exponent
\eqref{eq:rho} we obtain the following restriction for the effective mass
\begin{equation}\label{eq:discrmass}
\hat{m}^2=m_{\text{BF}}^2+\frac{(2n-1)^2}{4\ell^2},
\end{equation}
i.e., the related configurations involve only discrete values of mass with a
gap above the Breitenlohner-Freedman bound \eqref{eq:BF} for the first value,
making all of them physically acceptable. Additionally, each of these
discrete values presents a double degeneration allowing two solutions to
share the same mass value: the first characterized by a positive integer and the second by a nonpositive
one. These solutions must satisfy Eq.~\eqref{eq:massiveredef}, which becomes
\begin{equation}\label{eq:massiveredefn}
\left(\partial^{2}_{z\bar{z}}
+\frac{n}{z-\bar{z}}\partial_{\bar{z}}
-\frac{n}{z-\bar{z}}\partial_z\right)h=0,
\end{equation}
and can be straightforwardly read from the general solutions to the
Euler-Darboux equations we review in Appendix~\ref{App:ED}, with positive
[Eq.~\eqref{eq:solgen1}] and nonpositive [Eq.~\eqref{eq:solgen2}] integer parameters,
together with a reality condition. Consequently, the exact massive modes
$\mathscr{H}(u,z,\bar{z})$ satisfying Eq.~\eqref{eq:ComplexHDE} with the
discrete mass values \eqref{eq:discrmass} are necessarily given via the redefinition
\eqref{eq:H2h} by the profiles
\begin{align}
\mathscr{H}&=\left(\!\frac{z-\bar{z}}{2i}\!\right)^{n+1}
\mathfrak{u}(n,n)\nonumber\\
&=    \begin{cases}
2i\left(\!\dfrac{z-\bar{z}}{2i}\!\right)^{n+1}
\dfrac{\partial^{2(n-1)}}
{\partial{z}^{n-1}\partial{\bar{z}}^{n-1}}\left(
\dfrac{\varsigma+\overline{\varsigma}}{z-\bar{z}}
\right)\!, &  n>0,\\
2i\left(\!\dfrac{z-\bar{z}}{2i}\!\right)^{2-n}
\dfrac{\partial^{-2n}{}}
{\partial{z}^{-n}\partial{\bar{z}}^{-n}}\left(
\dfrac{\varsigma+\overline{\varsigma}}{z-\bar{z}}
\right)\!, &  n\leq0,
\rule[-4mm]{0pt}{11mm}
      \end{cases}
\end{align}
where $\varsigma=\varsigma(u,z)$ is a complex function that depends arbitrarily
on its arguments, but it is holomorphic in the complex wavefront
coordinate. Notice that we have made  appropriate choices for the coefficients in order to
have a manifestly real profile. For more general values of the mass we need
a different approach, explained below.

\subsubsection{Generic mass value: $\hat{m}^{2}>m_{\text{BF}}^{2}$}

Let us consider now a generic value for the mass in \eqref{eq:rho}. This
requires to integrate the Euler-Poisson-Darboux equation
\eqref{eq:massiveredef} for a generic value of its parameter. The involved
solution can be written from the general solutions \eqref{eq:gensolED} to the
Euler-Darboux equations we review in Appendix~C. Such solution is built as a
superposition of two linearly independent particular solutions which are
valid when their parameters take generic values such that
$\alpha+\beta\neq1$. For massive configurations $\alpha+\beta=2(\rho-1)$ and
the integral representation \eqref{eq:gensolED} applies for $\rho\neq3/2$,
i.e., the Breitenlohner-Freedman bound \eqref{eq:BF} cannot be saturated in
Eq.~\eqref{eq:rho}. Therefore, any exact massive mode above this bound,
$\hat{m}^{2}>m_{\text{BF}}^{2}$, is given by
\begin{align}\label{eq:integralH}
\mathscr{H}(u,z,\bar{z})={}&\left(\!\frac{z-\bar{z}}{2i}\!\right)^\rho
\mathfrak{u}_\text{g}(\rho-1,\rho-1)\nonumber\\
={}&
\left(\!\frac{z-\bar{z}}{2i}\!\right)^{3-\rho}\int_{0}^{1}
\varphi\!\left(u,z+(\bar{z}-z)t\right)\nonumber\\
&\qquad\qquad\qquad \times t^{1-\rho}(1-t)^{1-\rho}dt\nonumber\\
&+\left(\!\frac{z-\bar{z}}{2i}\!\right)^{\rho}\int_{0}^{1}
\psi\!\left(u,z+(\bar{z}-z)t\right)\nonumber\\
&\qquad\qquad\qquad \times t^{\rho-2}(1-t)^{\rho-2}dt,
\end{align}
where $\varphi$ and $\psi$ are two arbitrary real functions and $\rho$ is
given by Eq.~\eqref{eq:rho}. We remark that consistently the above expression is
real. The only effect of taking the complex conjugate
$\overline{\mathscr{H}}$ is that the complex argument of the arbitrary
functions changes by
\[
\overline{z+(\bar{z}-z)t}=z+(\bar{z}-z)(1-t),
\]
i.e.\ the interpolation parameter is reversed from $t$ to $1-t$. Using $1-t$
as the new integration parameter and the fact that the rest of each integral
is invariant under this reversing, the integral representation
\eqref{eq:integralH} remains intact and it is concluded that
\[
\overline{\mathscr{H}}=\mathscr{H}.
\]

As the last word, a corroboration of the fact that the solution
\eqref{eq:integralH} is not the most general one when the
Breitenlohner-Freedman bound $\rho=3/2$ ($\hat{m}^{2}=m_{\text{BF}}^{2}$) is approached
 resides in the fact that in such a limit we end up with a solution that possesses
a single arbitrary function, $\tilde{\varphi}=\varphi+\psi$, instead of two
as it should be for a second-order equation. In the following we explain how
to appropriately saturate this celebrated bound.

\subsubsection{Saturating the Breitenlohner-Freedman bound}

For $\alpha+\beta=2(\rho-1)=1$ the two solutions from which the superposition
\eqref{eq:integralH} is built is no longer linearly independent, see
Appendix~\ref{App:ED}. This occurs for $\rho=3/2$ or from \eqref{eq:rho} when the
Breitenlohner-Freedman bound is saturated
\begin{equation}\label{eq:BFstrtn}
\hat{m}^{2}=m_{\text{BF}}^{2}=-\frac9{4\ell^2}.
\end{equation}
However, intriguingly, the saturated solution can be obtained from the
generic one through a nontrivial limit exhibited in general for Euler-Darboux
equations in Appendix C. We start with the observation that the
generic solution \eqref{eq:integralH} can be rewritten as
\begin{align}
\mathscr{H}={}&\left(\!\frac{z-\bar{z}}{2i}\!\right)^{3-\rho}
\int_{0}^{1}dt\,t^{1-\rho}(1-t)^{1-\rho}
\Biggl(\tilde{\varphi}(z+(\bar{z}-z)t)\nonumber\\
&+\tilde{\psi}(z+(\bar{z}-z)t)
\frac{\left[\frac1{2i}(z-\bar{z})t(1-t)\right]^{2\rho-3}-1}
{2\rho-3}\Biggr),
\label{eq:rwrttn_gnrc}
\end{align}
where the arbitrary functions are properly redefined (see Appendix~\ref{App:ED}).
This expression is more appropriate for exploring the saturation of the bound
without losing generality. In fact, the most general solution for the
Breitenlohner-Freedman mass \eqref{eq:BFstrtn} is just obtained by taking the
limit $\hat{m}^{2}\rightarrow m_{\text{BF}}^{2}$ in the previous expression
\begin{align}
\mathscr{H}_\text{BF}={}&
\lim_{\hat{m}^{2}\rightarrow m_{\text{BF}}^{2}}\mathscr{H}\nonumber\\
={}&\lim_{\rho\rightarrow3/2}\left(\!\frac{z-\bar{z}}{2i}\!\right)^{\rho}
\mathfrak{u}_\text{g}(\rho-1,\rho-1)\nonumber\\
={}&\left(\!\frac{z-\bar{z}}{2i}\!\right)^{3/2}
\int_{0}^{1}\frac{dt}{\sqrt{t(1-t)}}
\Biggl[\tilde{\varphi}(z+(\bar{z}-z)t)\nonumber\\
&+\tilde{\psi}(z+(\bar{z}-z)t)
\ln\left(\frac{z-\bar{z}}{2i}t(1-t)\right)\Biggr].
\label{eq:BFgs}
\end{align}
This kind of logarithmic behavior emerges when massive configurations
approach the Breitenlohner-Freedman bound and is a characteristic of many
massive theories
\cite{AyonBeato:2004fq,AyonBeato:2005bm,AyonBeato:2005qq,AyonBeato:2009yq,
Bergshoeff:2014eca}.

\section{Matter coupling for AdS waves\label{Sec:Matter}}

A widely discussed topic in bigravity is how matter should be coupled to the
gravities. Since there is no experimental feedback, there are many
possibilities that seem (at least theoretically) consistent. One proposal to
democratically couple matter without reintroducing the Boulware-Deser ghost
is through the construction of an effective metric \cite{deRham:2014naa}
\begin{equation}\label{def:EffectiveMetric}
g_{\mu\nu}^\text{E}=\alpha^2 g_{\mu\nu}
+2\alpha\beta g_{\mu\lambda}{\gamma^{\lambda}}_{\nu}+\beta^2 f_{\mu\nu}.
\end{equation}
A remarkable feature of such a metric is that it is symmetric under the
simultaneous exchange of  the metrics $g\leftrightarrow f$ and couplings
$\alpha\leftrightarrow \beta$ for the usual case of interest where the
vierbein and metric formalism coincide. Hence, the resulting matter coupling
will be symmetric with respect to both metrics as it is, in fact, the vacuum
bigravity (\ref{eq:Bigravity}) itself. The full theory is then described by
the action
\begin{equation}\label{eq:EffectiveCoupling}
S[g,f,\text{Matter}]=S_{\text{bi}}[g,f]
+\int{d^4x\sqrt{-g^{\text{E}}}
\mathcal{L}_\text{M}(g_{\mu\nu}^\text{E},\ldots)},
\end{equation}
where the first contribution stands for the bigravity action
(\ref{eq:Bigravity}) and $\mathcal{L}_\text{M}$ is the matter Lagrangian
built with the effective metric \eqref{def:EffectiveMetric}.

The Einstein field equations now must include the contribution of the matter
sources
\begin{equation}\label{eq:EinsteinMatter}
G^{\mu}_{\,\nu}-\frac{m^2\kappa_{g}}{\kappa}V^{\mu}_{\,\nu}=
\kappa_{g}T^{\mu}_{\,\nu},\quad
\mathcal{G}^{\mu}_{\,\nu}
-\frac{m^2\kappa_{f}}{\kappa}\mathcal{V}^{\mu}_{\,\nu}=
\kappa_{f}\mathcal{T}^{\mu}_{\,\nu},
\end{equation}
where the energy-momentum tensors are defined by
\begin{align}
T_{\mu\nu} & \equiv
\frac{-2}{\sqrt{-g}}\frac{\delta
\left(\sqrt{-g^\text{E}}\mathcal{L}_\text{M}\right)}
{\delta g^{\mu\nu}}
=\frac{\sqrt{-g^\text{E}}}{\sqrt{-g}}T^\text{E}_{\rho\sigma}
\frac{\delta g_{\text{E}}^{\rho\sigma}}{\delta g^{\mu\nu}},\nonumber\\
\mathcal{T}_{\mu\nu} & \equiv
\frac{-2}{\sqrt{-f}}\frac{\delta
\left(\sqrt{-g^{\text{E}}}\mathcal{L}_\text{M}\right)}
{\delta f^{\mu\nu}}
=\frac{\sqrt{-g^{\text{E}}}}{\sqrt{-f}}T^\text{E}_{\rho\sigma}
\frac{\delta g_{\text{E}}^{\rho\sigma}}{\delta f^{\mu\nu}}.
\label{eq:EffectiveEMMatter}
\end{align}
In the second equalities we apply the chain rule to rewrite these tensors in
terms of the standard energy-momentum tensor with respect to the effective metric,
$T^\text{E}_{\mu\nu}$, which can be calculated as usual. However, the
variation of the effective metric requires the knowledge of the variation of
the square-root $\gamma$ matrix, which has a cumbersome structure
\cite{Bernard:2014bfa}. Some efforts to avoid such variation were made in
Ref.~\cite{Schmidt-May:2014xla} by contracting the Einstein equations
(\ref{eq:EinsteinMatter}) with the inverse of an appropriate Jacobian.
Notably, this is another difficulty that can be circumvented using the
linearizing properties of generalized Kerr-Schild transformations, which
turn the computation of such variation into a very easy task. Due to the form
of the $\gamma$ matrix (\ref{eq:gammaKerrSchild}) for generalized Kerr-Schild
transformations as (\ref{eq:AdSansatz}), the effective metric
(\ref{def:EffectiveMetric}) is reduced to the form
\begin{equation}\label{eq:EffectiveMetricKS}
g^{\text{E}}_{\mu\nu}=\left(\alpha+\beta C\right)
\left(\alpha g_{\mu\nu}+\frac{\beta}{C}f_{\mu\nu}\right),
\end{equation}
whose inverse is
\begin{equation}
g_{\text{E}}^{\mu\nu}=\frac{1}{\left(\alpha+\beta C\right)^3}
\left(\alpha g^{\mu\nu}+\beta C^3 f^{\mu\nu}\right).
 \label{eq:InvEffectiveMetricKS}
\end{equation}
This gives straightforwardly
\begin{equation}\label{eq:EffectiveMetricVariation}
\frac{\delta g_\text{E}^{\rho\sigma}}{\delta
g^{\mu\nu}}=\frac{\alpha}{\left(\alpha+\beta C\right)^3}
\delta^{\rho}_{(\mu}\delta^{\sigma}_{\nu)},\quad
\frac{\delta g_\text{E}^{\rho\sigma}}{\delta
f^{\mu\nu}}=\frac{\beta C^3}{\left(\alpha+\beta C\right)^3}
\delta^{\rho}_{(\mu}\delta^{\sigma}_{\nu)},
\end{equation}
and, consequently, the energy-momentum tensors contributing to each set of
Einstein equations only differ from the canonical one (calculated from the
effective metric) by constant factors
\begin{equation}\label{eq:E-Mtensors}
\frac1{\alpha}T_{\mu\nu}=
\frac{C}{\beta}\mathcal{T}_{\mu\nu}
=(\alpha+\beta C)T^{\text{E}}_{\mu\nu}.
\end{equation}

Another important consequence of the Kerr-Schild ansatz for the present
context is that, after fixing the effective cosmological constants as in the
vacuum \eqref{eq:AdSRelation}, the left-hand sides of both Einstein equations
\eqref{eq:EinsteinMatter} only have contributions along the null ray $k^\mu$,
see Eqs.~\eqref{eq:EinsteinEqs} and \eqref{eq:EinsteinAdSWaves}. This forces to
any matter supporting the AdS waves to behave as pure radiation (a null
pressureless fluid). The consequences of the resulting \emph{pure radiation
constraints} have been explored in standard gravity for scalar fields in
Refs.~\cite{AyonBeato:2005qq,AyonBeato:2006jf}.

Regarding the equations of motion for the matter fields, since they come from
matter variation of action \eqref{eq:EffectiveCoupling}, they necessarily have the
standard form but written in terms of the effective metric. We shall now consider two concrete examples of matter fields coupled to AdS waves in order to
test the previous considerations. We start with a massless free scalar field,
and follow with the study of the Maxwell field.

\subsection{Effective coupling to scalar fields}

Let us first consider the effective coupling to a massless free scalar field---the simplest matter one can think of. The Lagrangian only consists in the
kinetic term, but constructed with the effective metric
\begin{equation}\label{eq:SFLagrangian}
\mathcal{L}_\text{M}=-\frac{1}{2}g_\text{E}^{\mu\nu}
\partial_{\mu}\phi\partial_{\mu}\phi.
\end{equation}
This gives the standard energy-momentum tensor
\begin{equation}\label{eq:TEscalar}
T^{\text{E}}_{\mu\nu}=\partial_{\mu}\phi\partial_{\nu}\phi
-\frac{1}{2}g^\text{E}_{\mu\nu} g_\text{E}^{\rho\sigma}
\partial_{\rho}\phi\partial_{\sigma}\phi,
\end{equation}
together with the wave equation associated to the effective metric, which
rules the scalar field dynamics
\begin{equation}\label{eq:genKG}
\Box^{\text{E}}\phi=0.
\end{equation}

The scalar field is easily integrated from the emerging pure radiation
constraints, i.e., the vanishing of all the components of the energy-momentum
tensor except the one along the null ray, as is imposed by the Einstein equations
\cite{AyonBeato:2005qq,AyonBeato:2006jf}. Considering the
combination
\begin{align}\label{eq:PRCscalar}
0 &=2T^\text{E}_{uv}+\left(1-\frac{\alpha {F_1}
    +\beta{C}{F_{2}}}{\alpha+\beta{C}} \right)T^\text{E}_{vv} \nonumber \\
  &=(\partial_{v}\phi)^{2}+(\partial_{x}\phi)^{2}+(\partial_{y}\phi)^{2}.
\end{align}
the result is that the scalar field is an arbitrary function of the retarded
time
\begin{equation}\label{eq:SF}
\phi=\phi(u),
\end{equation}
which automatically satisfies the wave equation \eqref{eq:genKG}. This result
is exactly the same even if one attempts to promote the scalar field to be
self-interacting by adding a potential to the Lagrangian \eqref{eq:SFLagrangian}.
The outcome is that no potential is compatible with supporting an AdS wave
unless a non-minimal coupling to the involved metric is also incorporated
\cite{AyonBeato:2005qq,AyonBeato:2006jf}. Hence, considering only minimal
coupling to the effective metric, the more general situation is just that of
a massless free scalar field.

So far, we are left only with a contribution along the null ray in both
Einstein equations which gives rise to two inhomogeneous differential
equations for the profiles $F_{1}$ and $F_{2}$, that can be decoupled as in
the vacuum
\begin{subequations}\label{eq:scalarFHsys}
\begin{align}
y^{2}\Delta_\text{S}\mathscr{F}
&=\frac{2\kappa_{g}\kappa_{f}(\alpha+\beta{C})^{2}}{\kappa}\dot{\phi}^2y^{2}
\label{eq:EffFDES},\\
y^{2}\Delta_\text{S}\mathscr{H}-\ell^{2}\hat{m}^{2}\mathscr{H}
&=\frac{2(\beta\kappa_{f}-\alpha{C}\kappa_{g})(\alpha+\beta{C})}{C}
\dot{\phi}^2y^{2}.\label{eq:EffHDES}
\end{align}
\end{subequations}
Because we are dealing with an inhomogeneous linear system, its most general
solution is built by superposing the general solution to the homogeneous
(vacuum) version of the equations with any particular solution of the
inhomogeneous (with sources) ones
\begin{equation}\label{eq:gsi}
\mathscr{F}=\mathscr{F}^\text{h}+\mathscr{F}^\text{i}, \qquad
\mathscr{H}=\mathscr{H}^\text{h}+\mathscr{H}^\text{i}.
\end{equation}
The homogeneous version corresponds to the vacuum system (\ref{eq:FHsys}),
whose general solution was studied in detail in the last section and is given
by the Siklos solution (\ref{eq:solED1}) for the massless case
$\mathscr{F}^\text{h}$ and by the integral representation (\ref{eq:integralH})
for the massive one $\mathscr{H}^\text{h}$. In order to incorporate the
inhomogeneous contributions, due to the simple form of the scalar
inhomogeneities in Eqs.~\eqref{eq:scalarFHsys} it is enough to look for
particular solutions which are independent of the $x$ coordinate, then, the
above equations become inhomogeneous ordinary Euler equations for the
$y$ coordinate. The solutions are easily found and they are proportional to the power exhibited
at the inhomogeneity, except when that power resonates with one of the vacuum
power-law modes \eqref{eq:solsFH}.\footnote{We emphasize this is a genuine
resonance phenomenon. Notice we can change to a different coordinate
$y=\ell\exp(t/\ell)$ for which the ordinary Euler equations become linear
equations with constant coefficients, where the resonance phenomenon is
normally defined. There, the power-law solutions become exponentials and the
power exponents become mass-dependent frequencies. The resonance is
associated to the precise values of the parameters specifying the system and
is independent of the variables chosen to describe it.} The result is the
following
\begin{subequations}\label{eq:SFprofiles}
\begin{align}
\mathscr{F}^\text{i}&=
-\frac{\kappa_{g}\kappa_{f}(\alpha+\beta{C})^{2}}{\kappa}\dot{\phi}^2y^{2},
\rule[-4mm]{0pt}{4mm}
\label{eq:SF-Fp} \\
\mathscr{H}^\text{i}&=
\begin{cases}
-\dfrac{2(\beta\kappa_{f}-\alpha{C}\kappa_{g})(\alpha+\beta{C})}
{C(\hat{m}^2\ell^{2}+2)}\dot{\phi}^2y^2,
&\hat{m}^2\neq m_\text{sr}^2,\\
\dfrac{2(\beta\kappa_{f}-\alpha{C}\kappa_{g})(\alpha+\beta{C})}{C}
\dot{\phi}^2y^{2}\ln{\!\dfrac{y}{\ell}},
&\hat{m}^2=m_\text{sr}^2,\rule{0pt}{7mm}
\end{cases}
\label{eq:SF-Hp}
\end{align}
where the scalar source inhomogeneity enters in resonance with the vacuum
modes when the mass becomes
\begin{equation}\label{eq:srmass}
m_\text{sr}^2\equiv-\frac2{\ell^2}.
\end{equation}
\end{subequations}
What happens in this case is that one of the vacuum powers \eqref{eq:rho+-}
becomes $\rho_+=2$ and is equated by the scalar source value. We stress that
although the resulting scalar resonant mass is negative, it describes
physically admissible configurations above the Breitenlohner-Freedman bound
\[
m_\text{sr}^2=m_{\text{BF}}^{2}+\frac1{4\ell^2}.
\]
With respect to the inhomogeneous contributions \eqref{eq:SFprofiles}, nothing
special occurs for the rest of the masses, even for the
Breitenlohner-Freedman bound. In fact, the logarithmic behavior at
$\hat{m}^2=m_\text{sr}^2$ of \eqref{eq:SF-Hp} has a resonant origin and is
different from the one appearing at the vacuum for the multiplicity that
occurs when the Breitenlohner-Freedman bound is saturated.

Superposing these inhomogeneous contributions with those already studied for
the vacuum according to Eq.~\eqref{eq:gsi}, we obtain the most general form in
which a scalar field can support bigravity AdS waves. In what follows we
study the related, more complex problem for a Maxwell field.

\subsection{Effective coupling to Maxwell fields}

The following is another natural scenario in which bigravity could interact
with matter. We are interested now in how electromagnetic radiation fields
bend the AdS waves. For this purpose, consider the Maxwell Lagrangian
\begin{equation}\label{eq:EffectiveCouplingMaxwell}
\mathcal{L}_\text{M}=-\frac1{16\pi}
g_\text{E}^{\mu\rho}g_\text{E}^{\nu\sigma}F_{\mu\nu}F_{\rho\sigma},
\end{equation}
constructed with the effective metric, where the electromagnetic strength is
given as usual in terms of the vector potential,
$F_{\mu\nu}=2\partial_{[\mu}A_{\nu]}$. The standard electromagnetic
energy-momentum tensor resulting from varying with respect to the effective
metric is
\begin{equation}\label{eq:EffectiveMaxwellEM}
4\pi T^{\text{E}}_{\mu\nu}=
g_\text{E}^{\rho\sigma}F_{\mu\rho}F_{\nu\sigma}
-\frac14g^\text{E}_{\mu\nu}
g_\text{E}^{\gamma\rho}g_\text{E}^{\delta\sigma}
F_{\gamma\delta}F_{\rho\sigma}.
\end{equation}
Maxwell equations in terms of the effective metric are now obtained taking the variation
through the vector potential
\begin{equation}\label{eq:Maxwell}
\nabla_\text{E}^{\mu}
\left(g_\text{E}^{\nu\sigma}F_{\mu\nu}\right)=0.
\end{equation}

After a nontrivial gauge-fixing procedure described in
Appendix~\ref{App:Maxwell}, it is shown that the most general Maxwell potential
$A_\mu$ supporting AdS waves, i.e., that is compatible with the pure radiations
constraints, is proportional to the null rays $k_\mu$
\begin{equation}\label{eq:GaugeFixedMaxwellPot}
A=A_{u}(u,y,x)du.
\end{equation}
The Maxwell equations \eqref{eq:Maxwell} reduce to the harmonic equation
\begin{equation}
\Delta_\text{L}A_{u}=0,
 \label{eq:GaugedMaxwellEq}
\end{equation}
whose general solution is the real part of a general holomorphic function of
the complex wavefront coordinate $z=x+iy$
\begin{equation}\label{eq:PotMaxwellHarmonic}
A_{u}(u,y,x)=\partial_{z}a(u,z)
+\overline{\partial_{z}a(u,z)},
\end{equation}
where, for later convenience, we choose to write the holomorphic function as
the derivative of other holomorphic function $a(u,z)$. The only non-vanishing
components of the Faraday strength tensor are
\begin{equation}\label{eq:GaugedFaradayStrenght}
F_{ux}=-2\,\text{Re}\big(\partial^{2}_{z}a\big),
\qquad
F_{uy}=2\,\text{Im}\big(\partial^{2}_{z}a\big),
\end{equation}
and the effective energy-momentum tensor (\ref{eq:EffectiveMaxwellEM})
acquires the form of a pure radiation field [see \eqref{eq:EMMaxwell}]
\begin{align}
4\pi T^{\text{E}}_{\mu\nu}&=
\frac{y^4}{\ell^4}\frac{4|\partial^{2}_{z}a|^2}{(\alpha+\beta C)^2}
k_{\mu}k_{\nu}
\nonumber \\
{}&=\frac{y^4}{\ell^4}
\frac{\Delta_\text{L}\Delta_\text{L}(a\overline{a})}{4(\alpha+\beta C)^2}
k_{\mu}k_{\nu},
\label{eq:EffectiveEMTensorMaxwell}
\end{align}
which contributes to the equations for the AdS-wave profiles in the form of
inhomogeneities coming from the Maxwell field. The resulting equations can be
decoupled using the same combinations \eqref{def:SolvableCom}, just as in the
vacuum case. Thus, we get the inhomogeneous second order system
\begin{subequations}\label{eq:EffectiveFHsys}
\begin{align}
\Delta_\text{S}\mathscr{F}
&=\frac{\kappa_{g}\kappa_{f}}{8\pi\kappa\ell^2}
y^2\Delta_\text{L}\Delta_\text{L}(a\overline{a}), \label{eq:EffFDEM}\\
\Delta_\text{S}\mathscr{H}-\frac{\ell^{2}\hat{m}^2}{y^{2}}\mathscr{H}
&=\frac{(\beta\kappa_{f}-\alpha C\kappa_{g})}
{8\pi\ell^2C(\alpha+\beta C)}y^2
\Delta_\text{L}\Delta_\text{L}(a\overline{a}).\label{eq:EffHDEM}
\end{align}
\end{subequations}
The general solution is again represented by the superposition \eqref{eq:gsi}
with the same homogeneous contributions $\mathscr{F}^\text{h}$ and
$\mathscr{H}^\text{h}$ given by the vacuum configurations \eqref{eq:solED1} and
\eqref{eq:integralH}, respectively. The inhomogeneous contributions
$\mathscr{F}^\text{i}$ and $\mathscr{H}^\text{i}$ are due now to the more complex
Maxwell source, and understanding their behavior in the more general setting
requires a different treatment.

An interesting point to be noticed is that for both studied sources, if the
gravitational constants are tuned by the condition
\begin{equation}\label{eq:decoupleconst}
\alpha C\kappa_{g}-\beta\kappa_{f}=0,
\end{equation}
the massive profile $\mathscr{H}$ becomes decoupled from the sources and
behaves exactly as the vacuum configurations studied in
Sec.~\ref{Sec:GenSolSik}.

In order to fully understand the Maxwell contributions to bigravity AdS waves,
we shall proceed by gradually increasing the degree of difficulty. We start
by analyzing the massless sector, $\hat{m}^2=0$, which resembles that of
General Relativity with a Maxwell source originally studied by Siklos in
\cite{Siklos:1985}.

\subsubsection{Massless sector}

The inhomogeneous Siklos solution for the Maxwell source can be
straightforwardly read from the following fourth-order identity relating the
Siklos and Laplacian operators for any function $f$, see
Ref.~\cite{Siklos:1985},
\begin{equation}\label{eq:SiklosInIdentity}
\Delta_\text{S}\left(y^3\Delta_\text{L}f\right)
=y^2\Delta_\text{L}\Delta_\text{L}\left(yf\right).
\end{equation}
Applying the identity to the massless sector ($\hat{m}^2=0$) of the system
\eqref{eq:EffectiveFHsys} we obtain that the inhomogeneous contributions are
given by
\begin{subequations}
\begin{align}
\mathscr{F}^\text{i}&=\frac{\kappa_{g}\kappa_{f}}{8\pi\kappa\ell^2}y^{3}
\Delta_\text{L}\left(\frac{a\overline{a}}{y}\right),
\label{eq:FiM}\\
\mathscr{H}^\text{i}&=
\frac{(\beta\kappa_{f}-\alpha C\kappa_{g})}{8\pi\ell^2C(\alpha+\beta C)}y^{3}
\Delta_\text{L}\left(\frac{a\overline{a}}{y}\right), \quad \hat{m}^2=0.
\label{eq:HiMm=0}
\end{align}
\end{subequations}

Since the profile $\mathscr{F}$ describes a massless mode, their inhomogeneous
contribution \eqref{eq:FiM} remains the same even for more general values of
the mass. Hence, the most general massless configuration supported by
electromagnetism is given by the superposition profile \eqref{eq:gsi}
together with the Faraday strength following from the vector potential
\eqref{eq:GaugeFixedMaxwellPot}, which here take the explicit forms
\cite{Siklos:1985}
\begin{subequations}
\begin{align}
\mathscr{F}&=y^2\partial_{y}\!\left(\frac{\omega+\overline{\omega}}{y} \right)+
\frac{\kappa_{g}\kappa_{f}}{8\pi\kappa\ell^2}y^{3}
\Delta_\text{L}\left(\frac{a\overline{a}}{y}\right),\label{eq:Felectro}\\
F&=2\,\text{Re}(\partial^2_{z}a\,dz)\wedge du.\label{eq:Fmunu}
\end{align}
\end{subequations}
We know that the massless profiles are determined modulo residual symmetries
according to the transformation \eqref{eq:ResSymmetryFH}. This imposes that pairs of holomorphic
functions---determining the vacuum and electromagnetic contributions---that
are related under the transformation \eqref{eq:ressym} as
\begin{subequations}\label{eq:ResSymmE}
\begin{align}
\tilde{\omega}={}&(\lambda\mathsf{f})^2\biggl[\omega
-\frac{\kappa_{g}\kappa_{f}}{4\pi\kappa\ell^2}
(\overline{C}_1z+\overline{C}_0)a
\nonumber\\
&+\left(\frac{\kappa_{f}+C^{2}\kappa_{g}}2B_2
+\frac{\kappa_{g}\kappa_{f}}{8\pi\kappa\ell^2}|C_1|^2\right)z^{2}
\nonumber\\
&+\left(\frac{\kappa_{f}+C^{2}\kappa_{g}}2B_1
+\frac{\kappa_{g}\kappa_{f}}{8\pi\kappa\ell^2}
(C_0\overline{C}_1+\overline{C}_0C_1)
\right)z\nonumber\\
&+\frac{\kappa_{f}+C^{2}\kappa_{g}}2B_0
+\frac{\kappa_{g}\kappa_{f}}{8\pi\kappa\ell^2}|C_0|^2\biggr],\\
\tilde{a}={}&\lambda\mathsf{f}(a-C_1z-C_0),
\end{align}
\end{subequations}
all represent the same configuration. Here, $C_0=C_0(u)$ and $C_1=C_1(u)$ are
additional arbitrary complex functions of the retarded time which
parametrize the indetermination of the gauge field and which naturally
induce a generalization of the residual symmetry already known for the vacuum
\eqref{eq:ressym_omega}. This transformation was unveiled by Siklos in Ref. \cite{Siklos:1985}, without incorporating the diffeomorphic part. As he
emphasized, it is difficult to obtain due to the quadratic
contribution of the electromagnetic holomorphic dependence in the massless
geometric profile \eqref{eq:Felectro}. This exhausts the understanding of the massless excitations in presence of Maxwell sources.

The situation is different for the profile $\mathscr{H}$ which is massive in
nature; the inhomogeneous contribution \eqref{eq:HiMm=0} represents only a
particular case and the general solution for generic values of the mass with
Maxwell sources is a more difficult problem. There is no obvious
generalization of the identity \eqref{eq:SiklosInIdentity} which would
eventually allow to also derive a local expression for the massive solution.
However, the inhomogeneous solutions of any linear partial differential
equation generically allows an integral representation. Due to the connection
of the Siklos operator with the Euler-Darboux ones, we exploit this fact in
subsection~\ref{Subsubsec:Gms} to use a complexified version of the Riemann
method for hyperbolic equations, which we briefly review in Appendix~\ref{App:RG}.
But first, we attack the particular case of a Maxwell field whose strength is
homogeneous in the wavefront coordinates which ends up possessing an extra
symmetry; in this way the solution allows a local representation.

\subsubsection{Massive sector: wavefront-homogeneous Maxwell source}

Now we consider a particular example inspired by the fact that the
separable vacuum solutions studied in subsection.~\ref{Subsec:massive_sep},
modulo the use of residual symmetries, are AdS waves allowing as an additional
Killing vector the spatial translations $\partial_x$. Hence, we assume now
that the Maxwell field is compatible with such symmetry and consequently
$\partial_xF_{ux}=0=\partial_xF_{uy}$. Taking into account the general form
of the Faraday strength (\ref{eq:GaugedFaradayStrenght}), these conditions
imply
\begin{equation}\label{eq:MaxwellwithKilling}
\partial^{3}_{z}a=0 \quad \Longrightarrow \quad
a(u,z)=\frac12D_2z^2+D_1z+D_0,
\end{equation}
where $D_0=D_0(u)$, $D_1=D_1(u)$ and $D_2=D_2(u)$ are arbitrary complex functions
of the retarded time. Additionally, $D_0$ and $D_1$  can be eliminated by using the residual
transformation \eqref{eq:ResSymmE}. The resulting strength \eqref{eq:Fmunu}
is in general homogeneous in all of the wavefront coordinates. Keeping in mind
that the full electromagnetic contribution to the inhomogeneity of the
equations is now independent of the wavefront coordinate $x$, it is enough to
consider particular solution constructed in the same way, i.e.\
$\mathscr{F}^\text{i}=\mathscr{F}^\text{i}(u,y)$ and
$\mathscr{H}^\text{i}=\mathscr{H}^\text{i}(u,y)$. With this, the system
(\ref{eq:EffectiveFHsys}) becomes a pair of ordinary Euler equations
\begin{subequations}\label{eq:EffectiveFHsysKilling}
\begin{align}
(y^{2}\partial_{y}^{2}-2y\partial_{y})\mathscr{F} &
=\frac{2\kappa_{g}\kappa_{f}}{\pi\kappa\ell^2}
|D_2|^2y^4 \label{eq:EffFDEKilling},\\
(y^{2}\partial_{y}^{2}-2y\partial_{y}-\ell^{2}\hat{m}^{2})\mathscr{H} &
=\frac{2(\beta\kappa_{f}-\alpha C\kappa_{g})}
{\pi\ell^2C(\alpha+\beta C)}|D_2|^2y^4.
\label{eq:EffHDEKilling}
\end{align}
\end{subequations}
Consequently, for a Maxwell field homogeneous in the wavefront coordinates
\begin{subequations}\label{eq:ParticularSolFHsys}
\begin{equation}\label{eq:A(u,y)}
F=2\left[\text{Re}(D_2)dx-\text{Im}(D_2)dy\right]\wedge du,
\end{equation}
the inhomogeneous contributions in the decoupled AdS-wave profiles are given
by
\begin{align}
\mathscr{F}^\text{i}
&=\frac{\kappa_{g}\kappa_{f}}{2\pi\kappa\ell^2}|D_2|^2y^4,
\rule[-4mm]{0pt}{4mm}\\
\mathscr{H}^\text{i} &=
\begin{cases}
-\dfrac{2(\beta\kappa_{f}-\alpha C\kappa_{g})|D_2|^2}
{\pi\ell^2C(\alpha+\beta C)(\ell^2\hat{m}^2-4)}y^4, &
\hat{m}^2\neq {m}_\text{er}^2,\\
\dfrac{2(\beta\kappa_{f}-\alpha C\kappa_{g})|D_2|^2}
{5\pi\ell^2C(\alpha+\beta C)}y^4
\ln\!\dfrac{y}{\ell}, &
\hat{m}^2=m_\text{er}^2,
\rule{0pt}{7mm}
\end{cases}
\end{align}
where once again the inhomogeneity due to the electromagnetic source enters
in resonance with one of the vacuum power-law modes \eqref{eq:solsFH} when
the mass takes the value
\begin{equation}\label{eq:ermass}
m_\text{er}^2\equiv\frac4{\ell^2}.
\end{equation}
\end{subequations}
The electromagnetic resonance is produced because the power-law dictated by the electromagnetic source equals the vacuum power $\rho_+$ in Eq.~\eqref{eq:rho+-}, which takes the value $4$ at the above mass. The full solution is
obtained by superposing these inhomogeneous contributions with the vacuum ones
following  Eq.~\eqref{eq:gsi}, which describes all AdS waves of bigravity
supported by electromagnetic fields homogeneous in the wavefront coordinates.
The most general solutions with no additional restrictions on the Maxwell
sources are addressed in the next subsection.

\subsubsection{Massive sector: general Maxwell source\label{Subsubsec:Gms}}

As has been pointed throughout the paper, the Siklos operator and its massive
generalization can be expressed as Euler-Poisson-Darboux operators using
as independent variables the complex wavefront coordinate $z=x+iy$ and its
complex conjugate $\bar{z}$, together with the redefinition \eqref{eq:H2h}
for the massive profile. This allows to describe the behavior of massive
excitations in the presence of general Maxwell sources \eqref{eq:EffHDEM} by
means of an inhomogeneous Euler-Poisson-Darboux equation
\begin{equation}\label{eq:EPDmM}
E_{\rho-1,\rho-1}(h)=\frac{(\beta\kappa_{f}-\alpha C\kappa_{g})}
{32\pi\ell^2C(\alpha+\beta C)}\!
\left(\!\frac{z-\bar{z}}{2i}\!\right)^{2-\rho}\!\!
\Delta_\text{L}\Delta_\text{L}(a\overline{a}).
\end{equation}
The particular solutions of all inhomogeneous Euler-Darboux equations
\eqref{eq:InhEDe} can be obtained using the Riemann method reviewed in
Appendix~\ref{App:RG}. This allows to express the particular inhomogeneous
solution to massive excitations sourced by general Maxwell fields according
to Eqs.~\eqref{eq:H2h}, \eqref{eq:RGSol} and \eqref{eq:RGfuncED} as
\begin{align}
\mathscr{H}^\text{i}={}&\left(\!\frac{z-\bar{z}}{2i}\!\right)^\rho
\mathfrak{u}^\text{i}(\rho-1,\rho-1)\nonumber\\
={}&\left(\!\frac{z-\bar{z}}{2i}\!\right)^\rho\!
\int\!\!\!\int_{\Omega}\!{d\xi' d\eta'}f(u,\xi',\eta')\,
\mathfrak{R}(\xi',\eta';z,\bar{z})
\rule{0pt}{7mm}\nonumber\\
={}&\frac{(\beta\kappa_{f}-\alpha C\kappa_{g})}
{32\pi\ell^2C(\alpha+\beta C)}\!\left(\!\frac{z-\bar{z}}{2i}\!\right)^{\rho}
\rule{0pt}{7mm}\nonumber\\
&\times\frac12\!\left(
\int_{-\bar{z}}^z\!\! d\xi' \!\!\int_{-\xi'}^{\bar{z}}\!\! d\eta'\!+
\!\int_{-z}^{\bar{z}}\!\! d\eta'\!\!\int_{-\eta'}^z\!\! d\xi'\!\right)
\!\Delta_\text{L}'\Delta_\text{L}'(a\overline{a})
\rule{0pt}{7mm}\nonumber\\
&\times\left(\!\frac{\xi'-\eta'}{2i}\!\right)^{\rho}
\left(\frac{2i}{z-\eta'}\right)^{\rho-1}
\left(\frac{2i}{\xi'-\bar{z}}\right)^{\rho-1}\!
\rule{0pt}{7mm}\nonumber\\
&\times{_2F_1}\!\!\left(\rho-1,\rho-1;1;-\frac{(z-\xi')(\bar{z}-\eta')}
{(z-\eta')(\xi'-\bar{z})} \right)\!,
\rule{0pt}{7mm}
\label{eq:massInhEPDSolMax}
\end{align}
where ${_2F_1}$ describes the standard hypergeometric function.

We have used the double triangular integral representation \eqref{eq:PSolED}
with negative-inclined hypotenuse together with the election \eqref{eq:Hyperg} for the
solution, but the double representation with positive-inclined hypotenuse
\eqref{eq:PSolEDpi} is equally adequate. This is where these double
representations become relevant; despite its complexity, they significantly
ease the construction of  a real elliptic solution from the hyperbolic one.
In order to elucidate this far from obvious fact, we emphasize first the
following points: the nontrivial dependence of the exact massive excitation
in the complex wavefront coordinates emerges by evaluating the solution to
the standard hyperbolic inhomogeneous Euler-Darboux equation, as defined by
the double triangular integral \eqref{eq:PSolED}, in the following way
\begin{equation}\label{eq:H(z,zb)}
\mathscr{H}^\text{i}(u,z,\bar{z})\propto
\left.\mathfrak{u}^\text{i}(u,\xi,\eta)\right\rvert_{\xi=z,\eta=\bar{z}}.
\end{equation}
The inhomogeneity in Eq.~\eqref{eq:EPDmM} is a real function of the originally
real wavefront coordinates, which means that when it is written in terms of
the dummy variables inside the integral \eqref{eq:massInhEPDSolMax} obeys
that
\begin{align}
\overline{f(u,\xi',\eta')}&=
\overline{f\left(u,\frac{\xi'+\eta'}2,\frac{\xi'-\eta'}{2i}\right)}
\nonumber\\
&=f\left(u,\frac{\eta'+\xi'}2,\frac{\eta'-\xi'}{2i}\right)=f(u,\eta',\xi').
\label{eq:Realf}
\end{align}
Finally, the Riemann-Green function corresponding to the Euler-Darboux
equation \eqref{eq:RGfuncED} is real for real arguments and additionally
symmetrical under the simultaneous interchange of each pairs of variables,
i.e.\
\begin{equation}\label{eq:symmRG}
\mathfrak{R}(\eta',\xi';\eta,\xi)=\mathfrak{R}(\xi',\eta';\xi,\eta).
\end{equation}
We are now in a position to prove the reality of the solution
\eqref{eq:massInhEPDSolMax}; taking its complex conjugate we obtain
\begin{align}
\overline{\mathscr{H}^\text{i}}={}&\left(\!\frac{z-\bar{z}}{2i}\!\right)^\rho
\left.\overline{\mathfrak{u}^\text{i}(u,\xi,\eta)}
\right\rvert_{\xi=\bar{z},\eta=z}
\nonumber\\
={}&\left(\!\frac{z-\bar{z}}{2i}\!\right)^\rho\!
\frac12\!\left(
\int_{-z}^{\bar{z}}\!\! d\xi' \!\!\int_{-\xi'}^{z}\!\! d\eta'\!+
\!\int_{-\bar{z}}^{z}\!\! d\eta'\!\!\int_{-\eta'}^{\bar{z}}\!\! d\xi'\!\right)
\rule{0pt}{7mm}\nonumber\\
&\times\overline{f(u,\xi',\eta')}\,\mathfrak{R}(\xi',\eta';\bar{z},z)
\rule{0pt}{5mm}\nonumber\\
={}&\left(\!\frac{z-\bar{z}}{2i}\!\right)^\rho\!
\frac12\!\left(
\int_{-z}^{\bar{z}}\!\! d\eta'' \!\!\int_{-\eta''}^{z}\!\! d\xi''\!+
\!\int_{-\bar{z}}^{z}\!\! d\xi''\!\!\int_{-\xi''}^{\bar{z}}\!\! d\eta''\!\right)
\rule{0pt}{7mm}\nonumber\\
&\times
f(u,\xi''\!,\eta'')\,\mathfrak{R}(\xi''\!,\eta''\!;z,\bar{z})
\rule{0pt}{5mm}\nonumber\\
={}&\left(\!\frac{z-\bar{z}}{2i}\!\right)^\rho
\left.\mathfrak{u}^\text{i}(u,\xi,\eta)\right\rvert_{\xi=z,\eta=\bar{z}}
\rule{0pt}{7mm}\nonumber\\
={}&\mathscr{H}^\text{i},\rule{0pt}{4mm}
\end{align}
where the first and second equalities are self-explanatory, while in the third one
we use the interchanging properties \eqref{eq:Realf} and \eqref{eq:symmRG}
and later we reparametrize the dummy variables by
$(\xi',\eta')\mapsto(\xi''=\eta',\eta''=\xi')$. In terms of the new variables
both integrals are interchanged, which justifies the fourth equality. This
proof also works following the same steps if one uses the integral
representation for a triangle with positive-inclined hypotenuse
\eqref{eq:PSolEDpi}.

Another unavoidable exercise is to reconsider from the present perspective
the massless scenario $\hat{m}^2=0$ ($\rho=0$). This implies analyzing the
relation between the integral representation for the corresponding particular
inhomogeneous solution obtained from the Riemann method and the local one
already exhibited in Eq.~\eqref{eq:HiMm=0} that was originally provided by Siklos
\cite{Siklos:1985}. In fact, for $\rho=0$ the integral expression
\eqref{eq:massInhEPDSolMax} allows a double integration by parts. We find it
more easy to do the calculation using the analog triangular integral
representation with positive-inclined hypotenuse \eqref{eq:PSolEDpi}, which
gives
\begin{subequations}
\begin{align}
\mathscr{H}^\text{i}_\text{R}=\!{}&
-\frac{(\beta\kappa_{f}-\alpha C\kappa_{g})}
{32\pi\ell^2C(\alpha\!+\!\beta C)}
\frac12\!\!\left(
\int_{z}^{\bar{z}}\!\!d\xi'\!\!\int_{\xi'}^{\bar{z}}\!\!d\eta'\!+
\!\int_{z}^{\bar{z}}\!\!d\eta'\!\!\int_{{z}}^{\eta'}\!\!d\xi'\!\right)
\nonumber\\
&\times\Delta_\text{L}'\Delta_\text{L}'(a\overline{a})
\frac{(\xi'+\eta')(z+\bar{z})-2(\xi'\eta'+z\bar{z})}{(2i)^2}
\rule{0pt}{7mm}\nonumber\\
={}&\frac{(\beta\kappa_{f}-\alpha C\kappa_{g})}
{8\pi\ell^2C(\alpha\!+\!\beta C)}\text{Re}
\!\left(\int_{z}^{\bar{z}}\!\!d\xi'\!\!\int_{\xi'}^{\bar{z}}\!\!d\eta'
\partial^{2}_{\eta'\xi'}G(\xi',\eta';z,\bar{z})\!\!\right)
\rule{0pt}{7mm}\nonumber\\
={}&\frac{(\beta\kappa_{f}-\alpha C\kappa_{g})}
{8\pi\ell^2C(\alpha\!+\!\beta C)}
\Biggl[{}-G(z,\bar{z};z,\bar{z})
\rule{0pt}{7mm}\nonumber\\
&+\text{Re}\!\left(\!G(\bar{z},\bar{z};z,\bar{z})\!
-\!\!\int_{z}^{\bar{z}}\!\!\!d\xi'\!\!
\left.\bigl[\partial_{\xi'}G(\xi',\eta';z,\bar{z})\bigr]
\right\rvert_{\eta'=\xi'}\!\!\right)\!\Biggr]
\rule{0pt}{7mm}\nonumber\\
={}&\mathscr{H}^\text{i}_\text{S}+\mathscr{H}^\text{h},
\rule{0pt}{4mm}\label{eq:m0InhEPDSolMax}
\end{align}
where in the second equality we used the previously proven fact that the
second integral is the complex conjugate of the first. Additionally, the
integrand can be expressed as the mixed derivative of the following function
\begin{align}
G(\xi',\eta';z,\bar{z})\equiv{}&\Bigl\{
\left[(\xi'+\eta')(z+\bar{z})-2(\xi'\eta'+z\bar{z})\right]
\partial^2_{\xi'\eta'}\nonumber\\
&-(z+\bar{z}-2\xi')\partial_{\xi'}
-(z+\bar{z}-2\eta')\partial_{\eta'}\nonumber\\
&-2\Bigr\}a(\xi')\overline{a}(\eta'),
\end{align}
whose proper evaluation gives the Siklos inhomogeneous local solution as the
first contribution of the third equality
\begin{align}
\mathscr{H}^\text{i}_\text{S}=-{}&\frac{(\beta\kappa_{f}-\alpha C\kappa_{g})}
{8\pi\ell^2C(\alpha+\beta C)}G(z,\bar{z};z,\bar{z})\nonumber\\
={}&\frac{(\beta\kappa_{f}-\alpha C\kappa_{g})}{8\pi\ell^2C(\alpha+\beta C)}
\left(\!\frac{z-\bar{z}}{2i}\!\right)^{3}
\Delta_\text{L}\left(\frac{2i\,a\overline{a}}{z-\bar{z}}\right).
\end{align}
\end{subequations}
The remaining contribution defined as $\mathscr{H}^\text{h}$ is just a solution
to the homogeneous equation, which finally proves the compatibility between
the Riemann and Siklos particular inhomogeneous solutions in the massless
scenario.\\
Even though the result is more easy to obtain with the above representation, it is equally valid if the integration is performed over the triangle with a negative slope.
\subsection{Effective coupling to any matter source}

The Riemann method works so efficiently and in such a general fashion, that
one can go even one step further. Similar to the previously described AdS
waves supported by a Maxwell field, one can work in a more general setting
where the spacetime ripples of bigravity are caused by any kind of matter.
The first step is to consistently solve the related pure radiation
constraints, which by the proportionalities \eqref{eq:E-Mtensors} are all
expressed in terms of the energy-momentum tensor associated to the effective metric.
Using Bianchi identities this also entails solving the related field equations
that are naturally built on the effective background. If a nontrivial answer
emerges from this process,   it only remains to evaluate the surviving
contribution to the energy-momentum tensors along the retarded time and write
the single Einstein equation of each set. Considering as before the
decomposition \eqref{eq:H2h}, the dynamics of the decoupled exact excitations
are rigged now by the pair of inhomogeneous Euler-Poisson-Darboux equations
\begin{subequations}\label{eq:EPDmPRCmatt}
\begin{align}
E_{-1,-1}(\mathscr{F})&=\frac{\kappa_{g}\kappa_{f}(\alpha+\beta C)^2}{2\kappa}
T^\text{E}_{uu},\\
E_{\rho-1,\rho-1}(h)&=
\frac{(\alpha+\beta C)(\beta\kappa_{f}-\alpha C\kappa_{g})}{2C}\!
\left(\!\frac{z-\bar{z}}{2i}\!\right)^{-\rho}\!\!T^\text{E}_{uu}.
\end{align}
\end{subequations}
Thereby, in this more generic instance, they would be given by the following
expressions
\begin{widetext}
\begin{subequations}
\begin{align}
\mathscr{F}^\text{i}={}&\frac{\kappa_{g}\kappa_{f}(\alpha+\beta C)^2}{2\kappa}
\frac12\!\left(
\int_{-\bar{z}}^z\!\! d\xi' \!\!\int_{-\xi'}^{\bar{z}}\!\! d\eta'\!+
\!\int_{-z}^{\bar{z}}\!\! d\eta'\!\!\int_{-\eta'}^z\!\! d\xi'\!\right)
\times{T^\text{E}_{uu}}'\,
\frac{(\xi'+\eta')(z+\bar{z})-2(\xi'\eta'+z\bar{z})}{(\xi'-\eta')^2},
\rule{0pt}{7mm}\\
\mathscr{H}^{\text{i}}={}&\frac{(\alpha+\beta C)
(\beta\kappa_{f}-\alpha C\kappa_{g})}
{2C}\left(\!\frac{z-\bar{z}}{2i}\!\right)^{\rho}
\times\frac12\!\left(
\int_{-\bar{z}}^z\!\! d\xi' \!\!\int_{-\xi'}^{\bar{z}}\!\! d\eta'\!+
\!\int_{-z}^{\bar{z}}\!\! d\eta'\!\!\int_{-\eta'}^z\!\! d\xi'\!\right)
\!{T^\text{E}_{uu}}'
\rule{0pt}{7mm}\nonumber\\
&\times\left(\!\frac{\xi'-\eta'}{2i}\!\right)^{\rho-2}
\left(\frac{2i}{z-\eta'}\right)^{\rho-1}
\left(\frac{2i}{\xi'-\bar{z}}\right)^{\rho-1}\!
\times{_2F_1}\!\!\left(\rho-1,\rho-1;1;-\frac{(z-\xi')(\bar{z}-\eta')}
{(z-\eta')(\xi'-\bar{z})} \right)\!.
\rule{0pt}{7mm}
\end{align}
\end{subequations}
This completes the exhaustive scan of the AdS-wave configurations in bigravity.
\end{widetext}

\section{Conclusions \label{Sec:conclusions}}

In the preceding work we have tackled the problem of searching for exact
gravitational waves in the context of the ghost-free bimetric theory. Due to
the inherent presence of cosmological constants in this theory, we focused on
the Kundt class that is compatible with them, and in particular on the single example
where the involved nonexpanding null ray becomes a Killing symmetry: the AdS
waves. These configurations are additionally characterized by admitting a
Kerr-Schild representation. From previous results \cite{Ayon-Beato:2015qtt},
we are aware that this is an exceptional simplifying tool in the Hassan-Rosen
theory not only for exactly linearizing the Einstein equations as in General
Relativity, but also for the usually involved task of computing the interaction
between the metrics. The massive nature of bigravity manifests itself in a
very clear way, since the resulting wave equations for the profiles are
explicitly coupled by the mass term. Even though they can be decoupled
through redefinitions, one of the new profiles is inherently massive. We
learned how the physical d.o.f.---compatible with the symmetries
of the problem---are carried by the waves by looking for configurations that are
sum separable with respect to the coordinates defining the wavefront. In this
respect, another consequence of the presence of two metrics comes into play
even before turning on the mass. It is well known that spaces belonging to
the Kundt class possess residual symmetries \cite{Stephani:2003}, meaning
that some of the metric potentials can be gauged away with appropriate
transformations. In bigravity, we can only perform such transformations on
one of the metrics, implying that the d.o.f. that would be wiped out
in standard gravity remain present here, and actually propagate physical modes
of the theory.

Regarding the AdS-wave solutions, we can distinguish two cases in every
setting we studied. The massive profile depends on an effective mass that can
be turned off even if the flat-space Fierz-Pauli mass remains finite. This is
achieved by introducing a fine-tuning between the couplings, which produces
two copies of exact massless profiles. The general solution to this branch in
General Relativity was already reported by Siklos \cite{Siklos:1985}, whereas here we
explored the consequences of its observation in the sense that the
equation ruling the behavior of the wave profiles (now known as the Siklos
equation) turns out to be a particular complexified case of the Euler-Darboux
equations. This proves to be useful to understand the more complex scenarios
described below. The more general situation involving no fine-tuning
corresponds again to a massless profile, but it also includes a massive one. The
massive behavior is described by an extension of the previous Euler-Darboux
operators. One of the properties of the extended Euler-Darboux equations
consists in the possibility of eliminating nonderivative contributions
through a redefinition, returning to the standard Euler-Darboux equation. As
consequence, the massive part can be removed at the cost of generalizing the
parameters of the kinetic contributions. Once again, this allowed us to provide
the most general solution to the problem: a closed local form where the
Euler-Darboux parameters are integers, which is valid for a discrete family
of mass values, and an integral representation (originally due to Poisson)
for general mass values.

Since the wave profiles obey the (massive) wave equation over the AdS
spacetime, the well-known criterion of stability on AdS$_D$ of respecting the
Breitenlohner-Freedman bound $\hat{m}^2\geq-(D-1)^{2}/4\ell^2$, appears
repeatedly on the massive sector of the problem. With this in mind, it is interesting to inspect
the separable solutions. For $\hat{m}^{2}\neq-9/4\ell^2$,
we have profile solutions in powers of the conformal AdS coordinate $y$, which
include the standard Kaigorodov massless contribution \cite{Kaigorodov} that
cubically decays to infinity (located at $y=0$ in the AdS background in which
the modes are propagated) plus two other massive modes. For
$-9/4\ell^2<\hat{m}^{2}<0$, the massless Kaigorodov contribution becomes
subleading at infinity with respect to the decay of both massive
contributions. However, for $\hat{m}^{2}\geq0$ the leading massive
contribution no longer decays at infinity. In contrast, the other massive term decays and becomes subleading even
with respect to the massless
Kaigorodov one. Curiously enough, for the precise value
$\hat{m}^{2}=-9/4\ell^2$ (corresponding to the critical mass of the
Breitenlohner-Freedman bound) there is a degeneracy in the massive solutions, leading to logarithmic AdS waves---a
situation also observed in $2+1$ massive
gravities at critical points where they are supposed to be dual to
logarithmic conformal field theories
\cite{AyonBeato:2004fq,AyonBeato:2005bm,AyonBeato:2005qq,AyonBeato:2009yq,
Bergshoeff:2014eca}. This situation was replicated in the most general
solution for which the integral representation of the exact massive
excitation also acquires a logarithmic dependence when the
Breitenlohner-Freedman bound is saturated.

The natural step to follow was considering the wave dynamics in the presence
of matter; however, this question should be carefully examined in bigravity.
The problem of matter couplings has been widely discussed and the proposed
alternatives should always be concerned with not awaking again the undesired d.o.f. In this work we tested the approach
of introducing an effective
metric---a composite of the two original ones---constructed in such a way that
the whole theory is symmetric when replacing one original metric by the other
up to coupling redefinitions. In general, calculating the two involved energy-momentum tensors through this effective
metric is not an easy task, but we
found that even in this regard the Kerr-Schild ansatz provides another
significant simplification. By definition, the effective metric contains a
mixing term that is the product of one of the metrics with the coupling
square root matrix; this term is the origin of the difficulties when
performing metric variations to deduce the energy-momentum tensors.
Interestingly enough, under the adoption of Kerr-Schild forms this
interacting term is linearized and the effective metric results in a linear
superposition of both metrics. This leaves a straightforward calculation in
the variational problem and both energy-momentum tensors become proportional
to the standard one written in terms of the effective metric.

After unraveling how to calculate the energy-momentum tensors, another relevant
well-known issue when AdS waves are supported by matter sources is that their
Kerr-Schild structure is also replicated at the level of the Einstein equations.
In fact, their geometric left-hand side has only a contribution along
the ansatz null ray, which forces the involved matter to behave effectively
as pure radiation, i.e.,\ a null dust. This entails solving the resulting
\emph{pure radiation constraints} for the given source, and only when the
result is nontrivial can the AdS waves be supported by this kind of matter
\cite{AyonBeato:2005qq,AyonBeato:2006jf}. The first source analyzed was a
free massless scalar field. This situation is the simplest one, since the
outcome of solving the pure radiations constraints is that the most general
scalar field supporting an AdS wave is just an arbitrary function of the
retarded time. The resulting inhomogeneity added to the Siklos equation is a
simple quadratic dependence on the wavefront coordinate $y$. The
corresponding particular inhomogeneous part of the solutions obeys exactly
the same dependence since it can be assumed homogeneous in the other
wavefront coordinate, which reduces the Siklos equation to an Euler ordinary
equation. There is an exception when the mass value is such that the scalar
source enters in resonance with behavior characteristic of the vacuum; this
situation gives rise to logarithmic modes accompanied by the quadratic
resonance power.

The second type of matter source studied---the Maxwell field---is quite richer.
The most general self-gravitating vector potential that can be constructed
that is compatible with the pure radiation constraints has a single component along
the retarded time direction, which must be the real part of an arbitrary
holomorphic function of the complex coordinate formed with the wavefront
coordinates. This function also contains an undetermined dependence on the
retarded time. The resulting inhomogeneities added to the Siklos equation are
rather more involved now, since they possess a quadratic dependence on the
complex derivatives of this holomorphic function. The most general solution
to the inhomogeneous part for the massless cases was already presented by
Siklos himself \cite{Siklos:1985}. The tough part comes when dealing with the
sourced effectively massive equation. A preliminary simple case is to
consider a homogeneous Maxwell strength. This reduced the problem of finding a
particular inhomogeneous solution to solving again an ordinary Euler equation,
but this time with a quartic inhomogeneity in the conformal coordinate $y$.
The resulting solutions share the same $y$ dependence for all masses, except
for an electromagnetic resonant mass value for which the companion
coefficient develops the characteristic logarithmic behavior of resonance
phenomena. Describing the massive AdS-wave dynamics under the most general
Maxwell source is a highly nontrivial task. Here, once again, we
exploit the connection with the hyperbolic Euler-Darboux equations, whose inhomogeneous versions can be solved by means
of the Riemann
method. It consists in providing the general solution to a given initial
value problem by solving a related characteristic boundary value problem,
giving the so-called Riemann-Green function, which acts as the kernel in an
integral representation of the inhomogeneous solution. For the Euler-Darboux
equations it can be justified that their Riemann-Green functions are
determined in terms of hypergeometric functions \cite{Copson}. Adapting this
approach to our elliptic complexified paradigm allows us to provide the general
solution for the exact massive (massless) wave profiles not only when they
are supported by a Maxwell field, but also in the case when the source is
generalized to any matter consistent with the pure radiation constraints.
Whether the associated integral representation can be given a closed local
form or not depends on the specific value of the mass. For example, for zero
mass it can be shown that the particular Riemann solution consistently
reduces to the Siklos one modulo homogeneous solutions. A curious situation (common to all scenarios) is that the
matter can be decoupled from the massive
sector by means of a fine-tuning in the parameters defining the effective
metric.

To complete our work, we additionally studied the situation in which the
exact gravitational waves are propagated over flat spacetime. This involves
supplementary constraints on the bigravity coupling constants in order to get
rid of their naturally defined effective cosmological constants. Within the
Kundt class valid under this circumstance, the exceptional cases allowing
isometries are the well-known \emph{pp}-waves. Again, these exact waves were
decomposed into a massless excitation and a massive one. We analyzed in
detail the configurations that are sum separable with respect to the
wavefront coordinates. For the zero-mass case, we obtained two decoupled
massless profiles representing linearly polarized plane waves (standard in
General Relativity), plus linear and homogeneous terms in the wavefront
coordinates which are usually gauged from \emph{pp}-waves. However, here the
residual symmetries allow to remove the extra terms of only one of the
profiles; they remain untouched in the other and now describe the propagation
of the genuine physical d.o.f. of bigravity. In the properly massive
case, the decoupled solutions describe on the one hand the linearly polarized
plane wave of General Relativity for the massless mode, and on the other hand
a superposition of Yukawa exponential decays and growths in each wavefront
direction, corresponding to massive modes on flat spacetime.

Finally, we emphasize that not only are the methods introduced in this paper to deal with
exact gravitational waves in the presence of a cosmological constant original in the context of bigravity, but also
that there have been no similar studies even in General Relativity. We believe that these methods are useful beyond this
setting, since they teach us how to find the general solution to the
Klein-Gordon equation with sources on the AdS background, for configurations
which are only restricted to be invariant under a single light-cone
translation. It is not too risky to conjecture that these techniques could have
potential applications, for example, in the AdS/CFT context.

\begin{acknowledgments}
This research was supported by CONACyT Grant No. 175993. J.A.M.Z. was supported by Grant No. 243377 from CONACyT and
is partially funded by ``Convocatoria Max-Planck-CONACyT 2017 para estancias postdoctorales''. D.H.B. was supported by
Grant No. 243342 from CONACyT.
\end{acknowledgments}

\appendix

\section{\emph{pp}-Waves \label{App:pp-waves}}

As was briefly reviewed in Sec.~\ref{Sec:EGwaves}, the \emph{pp}-waves are
characterized within all of the exact gravitational waves by allowing a
covariantly constant null vector field (parallel rays) and the
requirement of propagating plane wavefronts, which is summarized in
the line element \cite{Brinkmann:1925fr}
\begin{equation}\label{eq:ppWaves}
ds^{2}=-F(u,\vec{x})du^{2}-2dudv+d\vec{x}^2,
\end{equation}
where $\vec{x}=(x,y)$ denotes a Euclidean vector. Consequently, the multiple
principal null direction is at the same time a Killing field $\partial_v$
and a gradient
\begin{equation}\label{eq:kpp}
k_{\mu}dx^{\mu}=-du.
\end{equation}
Hence, the \emph{pp}-waves are the simplest nontrivial example of a
Kerr-Schild transformation from Minkowski flat spacetime
\begin{equation}\label{eq:ppWaveK-S}
ds^2=ds_\text{M}^2-F \, k \otimes k,
\end{equation}
and again the null and geodesic character of $k$ warrants the linearization,
allowing the profile $F$ to obey the wave equation in General Relativity.
Thus, they are interpreted as exact gravitational waves propagating on the
flat background.

The \emph{pp}-wave metric \eqref{eq:ppWaves} is form invariant under the
family of transformations \cite{Stephani:2003}
\begin{subequations}\label{eq:ressympp}
\begin{align}
\tilde{u}&=\lambda\left(u+u_0\right),\nonumber\\
\vec{\tilde{x}}&=\tensor{\Lambda}\cdot\left(\vec{x}+\vec{P}\right),
\nonumber\\
\tilde{v}&=\lambda^{-1}\left[v+\dot{\vec{P}}\cdot\vec{x}+
           \frac12\int du\left(B_0+\dot{\vec{P}}^2\right)\right],
\nonumber\\
\tilde{F}&=\lambda^{-2}\left(F-\vec{B}_{1}\cdot\vec{x}-B_0
\right),
\end{align}
where $u_0$, $\lambda$, and the matrix $\tensor{\Lambda}\in SO(2)$ are
constants, $\vec{P}=\vec{P}(u)$ and $B_0=B_0(u)$ are arbitrary functions of
the retarded time, a dot denotes derivative with respect to $u$, and the
coefficients of the linear terms in the wavefront coordinates at the profile
transformation are determined from the above functions by
\begin{equation}\label{eq:ressympp2}
\vec{B}_{1}=2\ddot{\vec{P}}.
\end{equation}
\end{subequations}
Similar to the case of AdS waves, if the solution profile $F$ contains linear
terms in the wavefront coordinates $\vec{x}$ and/or a term which is only
function of the retarded time $u$, such contributions can be eliminated by
appropriate diffeomorphisms; see Ref. \cite{AyonBeato:2005bm}.

The starting point to explore the behavior of \emph{pp}-waves in bigravity is
a pair of Kerr-Schild ansatz, where the second metric is allowed to contain a
global conformal factor
\begin{align}
g_{\mu\nu}&=\eta_{\mu\nu}-F_{1}(u,\vec{x})k_{\mu}k_{\nu},\nonumber\\
f_{\mu\nu}&=C^{2}\left(\eta_{\mu\nu}-F_{2}(u,\vec{x})k_{\mu}k_{\nu}\right).
\label{eq:ppansatz}
\end{align}
As was previously argued in Sec.~\ref{Sec:AdSwaves}, the interaction square-root matrix has the form
(\ref{eq:gammaKerrSchild}) for any Kerr-Schild
ansatz and the interaction terms reduce to the same structure as the Einstein
equations (\ref{eq:EinsteinEqs}). The fact that the wavefronts of the
\emph{pp}-waves are planes implies there is no longer a
diagonal contribution in their Einstein tensors as those of the tensors \eqref{eq:EinsteinAdSWaves}; consequently,
the effective cosmological constants \eqref{eq:AdSRelation} do not appear in this case,
which imposes the constraints
\begin{equation}\label{eq:P1=0=P2}
P_{1}=0=P_{2}.
\end{equation}
The only nontrivial contributions are those of the off-diagonal components
along the null ray
\begin{subequations}\label{eq:ppF1F2sys}
\begin{align}
\left( \frac12\Delta_\text{L}F_{1}
+\frac{C\kappa_{g}m^{2}P_{0}}{2\kappa}(F_{2}-F_{1})\right)
k_\mu k_\nu&=0,
\label{eq:ppDEF1}\\
\left( \frac12\Delta_\text{L}F_{2}
-\frac{\kappa_{f}m^{2}P_{0}}{2C\kappa}(F_{2}-F_{1})\right)
k_\mu k_\nu&=0.
\label{eq:ppDEF2}
\end{align}
\end{subequations}
This system becomes decoupled for the same combinations $\mathscr{F}$ and
$\mathscr{H}$ decoupling the AdS waves \eqref{def:SolvableCom}, which yields
the following equations defining again exact massless and massive excitations
propagating now in flat spacetime
\begin{subequations}\label{eq:ppFHsys}
\begin{align}
\Delta_\text{L}\mathscr{F}&=0, \label{eq:ppDEF}\\
\Delta_\text{L}\mathscr{H}-\hat{m}^{2}\mathscr{H}&=0. \label{eq:ppDEH}
\end{align}
\end{subequations}
The effective mass appearing in the later Klein-Gordon equation is the same as that already defined in Eq.
(\ref{def:meff}). These decoupled profiles are determined
up to the residual symmetries of the \emph{pp}-waves \eqref{eq:ressympp},
since they change according to
\begin{subequations}\label{eq:ResSymmetryFHpp}
\begin{align}
\tilde{\mathscr{F}}&=\lambda^{-2}
\left[\mathscr{F}-(\kappa_{f}+C^{2}\kappa_{g})
(\vec{B}_{1}\cdot\vec{x}+B_{0})\right],\label{eq:ResSymmetryFpp}\\
\tilde{\mathscr{H}}&=\lambda^{-2}\mathscr{H}.\label{eq:ResSymmetryHpp}
\end{align}
\end{subequations}
Similar to the case of AdS waves, only the massless profile inherits the
indeterminacy and the massive one is almost preserved modulo a trivial
scaling.

The decoupled exact massless profile satisfies the wave equation which, due
to the Killing vector $\partial_v$, becomes just the harmonic equation
\eqref{eq:ppDEF} with general solution
\begin{equation}\label{eq:HarmonicSol}
\mathscr{F}(u,\vec{x})=G(u,z)+\overline{G(u,z)},
\end{equation}
where $G(u,z)$ is any function depending arbitrarily on the retarded time, but
that is holomorphic in the complex wavefront coordinate $z=x+iy$. A
particularly interesting case is the fully massless one $\hat{m}=0$
($P_0=0$). Looking back to Eq. (\ref{eq:ppF1F2sys}), both of the original profiles are
harmonic
\begin{equation}\label{eq:HarmonicF1F2}
\Delta_\text{L}F_{1}=0=\Delta_\text{L}F_{2},
\end{equation}
and thus are described as in Eq. \eqref{eq:HarmonicSol}. Even with this quite
general solution, it is useful to explore the particular cases that are sum separable with respect to the wavefront
coordinates since they also bring here a
clear decomposition in the prevailing modes of the theory after removing the
unphysical contributions by means of the \emph{pp}-waves residual symmetry
(\ref{eq:ressympp}). Thus, we will search for the most general solutions of the
form $F_1(u,\vec{x})=X_{1}(u,x)+Y_{1}(u,y)$ and
$F_2(u,\vec{x})=X_{2}(u,x)+Y_{2}(u,y)$. Such solutions are
\begin{subequations}\label{eq:solppF1F2m=0}
\begin{align}
F_{1}(u,\vec{x})&=f_{2}(u)(x^{2}-y^{2}), \\
F_{2}(u,\vec{x})&=h_{2}(u)(x^{2}-y^{2})+\vec{h}_{1}(u)\cdot\vec{x}+h_{0}(u),
\end{align}
\end{subequations}
where all functions of the retarded time are arbitrary and we have eliminated
the linear and homogeneous terms in the wavefront coordinates appearing in
the first profile by using the residual symmetry \eqref{eq:ressympp}. The
quadratic contributions to both profiles are just the sum-separable part of
the well-known plane waves of General Relativity \cite{Stephani:2003}, which
corresponds to the linearly polarized ones \cite{Griffiths:1991zp}.

Applying the same analysis to the general massive case $\hat{m}\neq 0$, the
sum-separable solutions of Eq. (\ref{eq:ppFHsys}) are given by
\begin{subequations}\label{eq:solppHFa}
\begin{align}
\mathscr{F}(u,\vec{x}) &= f_{2}(u)(x^{2}-y^{2}), \\
\mathscr{H}(u,\vec{x}) &= \vec{h}_{+}(u)\cdot{e}^{\hat{m}\vec{x}}
+\vec{h}_{-}(u)\cdot{e}^{-\hat{m}\vec{x}}.
\end{align}
\end{subequations}
where $e^{\pm\hat{m}\vec{x}}=(e^{\pm\hat{m}x},e^{\pm\hat{m}y})$ denote Euclidean vectors and the residual symmetry
\eqref{eq:ResSymmetryFHpp} allows
us to get rid of the unphysical contributions in the massless profile. Here,
the massless profile again corresponds to the linearly polarized plane-wave
contribution of General Relativity, and the massive one describes a
superposition of Yukawa exponential decays and growths in each wavefront
direction, corresponding to standard massive modes in flat spacetime. From
the inversions \eqref{def:F1F2}, one gets the solutions for the original
profiles
\begin{subequations}
\begin{align}
F_{1}(u,\vec{x})={}&\frac1{\kappa_{f}+C^{2}\kappa_{g}}
\Bigl[\, \kappa f_{2}(u)(x^{2}-y^{2})\nonumber\\
&-C^2\kappa_{g}\left(\vec{h}_{+}(u)\cdot{e}^{\hat{m}\vec{x}}
+\vec{h}_{-}(u)\cdot{e}^{-\hat{m}\vec{x}}\right)\Bigr],\\
F_{2}(u,\vec{x})={}&\frac1{\kappa_{f}+C^{2}\kappa_{g}}
\Bigl[\, \kappa f_{2}(u)(x^{2}-y^{2})\nonumber\\
&+\kappa_{f}\left(\vec{h}_{+}(u)\cdot{e}^{\hat{m}\vec{x}}
+\vec{h}_{-}(u)\cdot{e}^{-\hat{m}\vec{x}}\right)\Bigr].
\end{align}
\end{subequations}
Let us note that it is not straightforward to obtain the $\hat{m}=0$ solution
from the last outcome. We end here the revision of bigravity \emph{pp}-waves.

\section{Most general vector potential supporting AdS waves
\label{App:Maxwell}}

As emphasized in Sec.~\ref{Sec:Matter}, the fact that Einstein tensors
have the structure \eqref{eq:EinsteinAdSWaves} for AdS waves
(\ref{eq:Siklos}) fixes the effective cosmological constants coming from the
interaction contributions as in Eq. \eqref{eq:AdSRelation}, which in turn implies
that both Einstein equations \eqref{eq:EinsteinMatter} establish the
vanishing of all of the components of the energy-momentum tensors
(\ref{eq:E-Mtensors}) that are not exclusively in the direction of the null
ray $k^\mu$. In particular, for a Maxwell field the following constraints
must be satisfied
\begin{subequations}\label{eq:PureRadiationConst}
\begin{align}
4\pi T_{vv}^\text{E}=
\frac{y^2}{l^2}\frac{F_{vx}^2+F_{vy}^2}{(\alpha+C\beta)^2} &=0,\\
4\pi\left(2 T_{uv}^\text{E}-
\frac{\alpha F_1+C\beta F_2}{\alpha+C\beta}T_{vv}^\text{E}\right)=
\frac{y^2}{l^2}\frac{F_{vu}^2+F_{xy}^2}{(\alpha+C\beta)^2} & =0.
\end{align}
\end{subequations}
These imply that the involved strength components vanish
\begin{equation}\label{eq:F's=0}
F_{vx}=F_{vy}=F_{vu}=F_{xy}=0,
\end{equation}
and that the energy-momentum tensor (\ref{eq:EffectiveMaxwellEM}) acquires
the form of a pure radiation field along the null ray
\begin{equation}\label{eq:EMMaxwell}
4\pi T_{\mu\nu}^\text{E}=\frac{y^4}{l^4}
\frac{F_{xu}^{2}+F_{yu}^{2}}{(\alpha+\beta C)^2}k_{\mu}k_{\nu}.
\end{equation}

Let us start by analyzing the consequences of the first three pure radiation
constraints \eqref{eq:F's=0}
\begin{equation}\label{eq:PRCsetA}
F_{va}=\partial_{v}A_a-\partial_aA_{v}=0,
\end{equation}
with $x^a\neq v$, which can be integrated as
\begin{equation}\label{eq:GaugeFixingA}
A_a=\partial_a\int{A_{v}dv}+\hat{A}_a(u,y,x),
\end{equation}
where $\hat{A}_a$ are integration functions which are independent of the null
ray parameter $v$. Consequently, modulo the gauge transformation
\begin{equation}\label{eq:GaugeFixingB}
A_\alpha\rightarrow
\tilde{A}_{\alpha}=A_{\alpha}-\partial_{\alpha}\int{A_{v}}dv,
\end{equation}
we end with a gauge where
\begin{equation}\label{eq:GaugeFixingC}
\tilde{A}_{a}=\hat{A}_{a}(u,y,x), \qquad \tilde{A}_{v}=0.
\end{equation}
Implementing now the last pure radiation constraint \eqref{eq:F's=0} in the
above gauge
\begin{equation}
F_{xy}=\partial_{x}\tilde{A}_{y}-\partial_{y}\tilde{A}_{x}=0,
\end{equation}
lead us to
\begin{equation}\label{eq:GaugeFixingD}
\tilde{A}_y=\partial_{y}\int{\tilde{A}_{x}dx}+\hat{\hat{A}}_{y}(u,y).
\end{equation}
The gauge \eqref{eq:GaugeFixingC} is still preserved by residual gauge
transformations, from which we choose
\begin{equation}\label{eq:GaugeFixingE}
\tilde{A}_{\alpha}\rightarrow
\tilde{\tilde{A}}_{\alpha}=\tilde{A}_{\alpha}
-\partial_{\alpha}\int{\tilde{A}_x dx},
\end{equation}
which allows a further fixing after using Eq. \eqref{eq:GaugeFixingD}
\begin{equation}\label{eq:GaugeCompatibility}
\tilde{\tilde{A}}_{u}=\tilde{\tilde{A}}_{u}(u,y,x),\quad
\tilde{\tilde{A}}_{v}=0=\tilde{\tilde{A}}_{x},\quad
\tilde{\tilde{A}}_{y}=\hat{\hat{A}}_{y}(u,y).
\end{equation}
Hence, satisfying the pure radiation constraints entails the existence of a
gauge where the vector potential can be expressed as
\begin{equation}\label{eq:GaugedMaxwellField}
A=A_{u}(u,y,x)du+A_{y}(u,y)dy.
\end{equation}

It remains to explore the repercussions of this gauge fixing on the Maxwell
equations (\ref{eq:Maxwell}), which become
\begin{equation}\label{eq:MaxwellEqnA}
\Delta_\text{L}A_{u}=\partial_{y}\partial_{u}A_{y}.
\end{equation}
This can be understood as an inhomogeneous linear partial differential
equation for the retarded time component $A_u$, where the inhomogeneity is
determine by the spatial component $A_y$. The most general solution is a
superposition of the kind \eqref{eq:gsi} for the potential
$A_{u}=A_{u}^\text{h}+A_{u}^\text{i}$. The contribution $A_{u}^\text{h}$ must
be harmonic since it represents the general solution to the related
homogeneous equation, and can be written as the real part of a general
holomorphic function in the complex wavefront coordinate $z=x+iy$
\begin{equation}\label{eq:PotMaxwellHarmonic2}
A_{u}^\text{h}(u,y,x)=\partial_{z}a(u,z)
+\overline{\partial_{z}a(u,z)}.
\end{equation}
Here, for further applications it is convenient to write the holomorphic
function as the derivative of another holomorphic function $a(u,z)$. Regarding
the inhomogeneous contribution $A_{u}^\text{i}$, since only a particular
solution to the inhomogeneous equation is needed, we can assume that it does
not depend on the spatial coordinate $x$, i.e.,\
$A_{u}^\text{i}=A_{u}^\text{i}(u,y)$. Thus, the inhomogeneous equation
\eqref{eq:MaxwellEqnA} is reduced to
\begin{equation}\label{eq:MaxwellParticular}
\partial_{y}(\partial_{y}A_{u}^\text{i}-\partial_{u}A_{y})=0,
\end{equation}
and is integrated as
\begin{equation}\label{eq:MaxwellParticularA}
A_{u}^\text{i}(u,y)=\partial_{u}\int{A_{y}(u,y)dy}+K_1(u)y+K_2(u).
\end{equation}
The simplest particular solution is obtained by choosing $K_1=0=K_2$.
Consequently, the most general solution to the Maxwell equation
\eqref{eq:MaxwellEqnA} is
\begin{equation}\label{eq:MaxwellSol}
A_{u}(u,y,x)=\partial_{z}a(u,z)+\overline{\partial_{z}a(u,z)}
+\partial_{u}\int{A_{y}(u,y)dy}.
\end{equation}
However, the last expression suggests a third gauge transformation
\begin{equation}\label{eq:GaugedMaxwellSol}
A_{\alpha}\rightarrow
\tilde{\tilde{\tilde{A}}}_{\alpha}=
A_{\alpha}-\partial_{\alpha}\int{A_{y}(u,y)dy},
\end{equation}
which further restrict the gauge to its final form
\begin{equation}\label{eq:GaugeCompatibilityB}
\tilde{\tilde{\tilde{A}}}_{u}=A_{u}^\text{h}(u,y,x),\quad
\tilde{\tilde{\tilde{A}}}_{v}=
\tilde{\tilde{\tilde{A}}}_{x}=
\tilde{\tilde{\tilde{A}}}_{y}=0,
\end{equation}
i.e.,\ all components but $A_{u}$ are pure gauge. Finally, after solving the
pure radiation constraints and the Maxwell equations, the vector potential is
completely determined. This allows us to conclude that there exists a gauge
where the most general self-gravitating Maxwell potential supporting AdS
waves can be written as
\begin{equation}\label{eq:FinalMaxwellPotential}
A=A_{u}(u,y,x)du=
\left[\partial_{z}a(u,z)+\overline{\partial_{z}a(u,z)}\right]du.
\end{equation}
The first equality precisely corresponds to the generalization of the
Kerr-Schild ansatz to include a Maxwell field, where the vector potential is
chosen proportional to the null vector $A_\mu \propto k_\mu$ [see, for example,
Ref. \cite{Ayon-Beato:2015qtt} where the Kerr-Newmann-(A)dS black hole was derived
for bigravity following this approach, inspired by the results of
Ref. \cite{Ayon-Beato:2015nvz} for General Relativity].

\section{The Euler-Darboux equations \label{App:ED}}

In this appendix we review the main features of solving the so-called
Euler-Darboux equations \cite{Euler:1770,Poisson:1823,Darboux:1899}, since the
dynamics of the four-dimensional AdS waves (either massless or massive) can
be reduced to their study. Our discussion mainly follows
Ref.~\cite{Koshlyakov}, but highlights several aspects not found there. The
Euler-Darboux equations are defined as
\begin{equation}\label{eq:ede}
E_{\alpha,\beta}(\mathfrak{u})\equiv
\left(\partial^2_{\xi\eta}+\frac{\alpha}{\xi-\eta}\partial_{\eta}
-\frac{\beta}{\xi-\eta}\partial_{\xi}\right)\mathfrak{u}=0,
\end{equation}
where, according to the Darboux notation \cite{Darboux:1899}, $\alpha$ and
$\beta$ are real parameters labeling not only the above operators, but also
any solution to the previous equation by $\mathfrak{u}(\alpha,\beta)$. For
likewise real variables $\xi$ and $\eta$, it is the prototype of a
hyperbolic equation \cite{Koshlyakov}. The case of coinciding parameters is
named the Euler-Poisson-Darboux equation
\cite{Euler:1770,Poisson:1823,Darboux:1899}. When $\alpha+\beta\neq1$ one can
introduce a new function using
\begin{equation}
\mathfrak{u}=(\xi-\eta)^{1-\alpha-\beta}\mathfrak{v},\nonumber
\end{equation}
and by replacing it back into Eq. \eqref{eq:ede} it is easy to see that $\mathfrak{v}$
must indeed satisfy another Euler-Darboux equation with parameters $1-\beta$
and $1-\alpha$, i.e.
\begin{equation*}
E_{1-\beta,1-\alpha}(\mathfrak{v})=0.
\end{equation*}
Namely, there exists a relation between the solutions of the Euler-Darboux
equation with parameters $\alpha, \beta$ and those with parameters $1-\beta,
1-\alpha$
\begin{equation}\label{eq:firstrel}
\mathfrak{u}(\alpha,\beta)=(\xi-\eta)^{1-\alpha-\beta}
\mathfrak{u}(1-\beta,1-\alpha).
\end{equation}
The previous relation is useful, for example, to straightforwardly find the
most general solution to the equation $E_{1,1}(\mathfrak{u})=0$ as
\begin{equation}\label{eq:u11}
\mathfrak{u}(1,1)=\frac{\mathfrak{u}(0,0)}{\xi-\eta}
=\frac{f(\xi)+g(\eta)}{\xi-\eta},
\end{equation}
where $\mathfrak{u}(0,0)$ is the solution to the two-dimensional wave
equation in light-cone coordinates, $E_{0,0}(\mathfrak{u})=0$, which in
general is separable into left and right movers encoded in the arbitrary
functions $f$ and $g$.

The general solutions for other integer values of the parameters can also be
easily expressed. In order to accomplish this, another important feature to
realize is that the different solutions are related not only algebraically [as
in Eq. \eqref{eq:firstrel}], but also through differentiation. For example, if we
know a solution of $E_{\alpha,\beta}(\mathfrak{u})=0$, their derivatives
\begin{equation}\label{eq:secondrel}
\mathfrak{u}(\alpha+m-1,\beta+n-1)=
\frac{\partial^{m+n-2}\mathfrak{u}(\alpha,\beta)}
{\partial\xi^{m-1}\partial\eta^{n-1}},
\end{equation}
are also solutions, in this case to
$E_{\alpha+m-1,\beta+n-1}(\mathfrak{u})=0$. In particular, by evaluating
$\alpha=1=\beta$, we have that
\begin{align}\label{eq:solgen1}
\mathfrak{u}(m,n)&=\frac{\partial^{m+n-2}\mathfrak{u}(1,1)}
{\partial\xi^{m-1}\partial\eta^{n-1}}\nonumber\\
&=\frac{\partial^{m+n-2}{}}{\partial\xi^{m-1}\partial\eta^{n-1}}
\left(\frac{f(\xi)+g(\eta)}{\xi-\eta}\right),
\end{align}
i.e.,\ the general solution to the equation $E_{m,n}(\mathfrak{u})=0$ where
$m$ and $n$ are positive integers can be obtained from multiple
differentiations of Eq. \eqref{eq:u11}. These solutions [in particular, their
Euler-Poisson-Darboux version $\mathfrak{u}(m,m)$] have been connected with
the characterization of the radial time-dependent part of the spherical
harmonic decomposition of the scalar wave equation near the spatial infinity
of the Schwarzschild black hole \cite{Schmidt:1978tk}.

The general solution for nonpositive integers can be obtained in a similar
way. It is enough to use the form of Eq. \eqref{eq:secondrel} on the left-hand
side of the general algebraic relation \eqref{eq:firstrel} to arrive at a new
differential identity between solutions
\begin{equation}\label{eq:thirdrel}
\mathfrak{u}(2-\beta-n,2-\alpha-m)
=(\xi-\eta)^{\alpha+\beta+m+n-3}
\frac{\partial^{m+n-2}\mathfrak{u}(\alpha,\beta)}
{\partial\xi^{m-1}\partial\eta^{n-1}}.
\end{equation}
Now, by again evaluating $\alpha=1=\beta$ and redefining the integers as
$m'=n-1$ and $n'=m-1$, we obtain
\begin{align}\label{eq:solgen2}
\mathfrak{u}(-m',-n')&=(\xi-\eta)^{m'+n'+1}
\frac{\partial^{m'+n'}\mathfrak{u}(1,1)}{\partial\xi^{n'}\partial\eta^{m'}}
\nonumber\\
&=(\xi-\eta)^{m'+n'+1}
\frac{\partial^{m'+n'}}{\partial\xi^{n'}\partial\eta^{m'}}
\left(\frac{f(\xi)+g(\eta)}{\xi-\eta}\right).
\end{align}
Hence, the general solutions to the Euler-Darboux equations where the
parameters take nonpositive integer values are once more generated from
Eq. \eqref{eq:u11} by differentiation. As is explained in
Sec.~\ref{Sec:GenSolSik}, the Euler-Poisson-Darboux subcase
$\mathfrak{u}(-1,-1)$ is just the solution of the Siklos equation
characterizing the AdS waves of General Relativity in presence of a negative
cosmological constant \cite{Siklos:1985}.

We now consider the case where the parameters $\alpha$ and
$\beta$ appearing in Eq. \eqref{eq:ede} are real numbers. The first step
is to look for particular separable solutions of the form
\begin{equation}
\mathfrak{u}_\text{p}(\alpha,\beta)=X(\xi)Y(\eta).
\end{equation}
By inserting this into the Euler-Darboux equation \eqref{eq:ede}, we obtain the
following solution
\begin{equation}\label{eq:upar1}
\mathfrak{u}_\text{p1}(\alpha,\beta)=(\xi-a)^{-\alpha}(\eta-a)^{-\beta},
\end{equation}
where $a$ is the separation constant. Moreover, we can use the relation
\eqref{eq:firstrel}, when $\alpha+\beta\neq1$, to construct another
particular solution of $E_{\alpha,\beta}(\mathfrak{u})=0$ as follows
\begin{equation}\label{eq:upar2}
\mathfrak{u}_\text{p2}(\alpha,\beta)=(\xi-\eta)^{1-\alpha-\beta}
(\xi-a)^{\beta-1}(\eta-a)^{\alpha-1}.
\end{equation}
The previous particular solutions are the seeds to construct the general
solution to the linear Euler-Darboux equation \eqref{eq:ede} as a
superposition
\begin{align}
\!\!\!\!\mathfrak{u}_\text{g}(\alpha,\beta)\!=\!{}&
\!\int_{\eta}^{\xi}\!\!\varphi(a)(\xi-a)^{-\alpha}(a-\eta)^{-\beta} da\nonumber\\
&\!+\!(\xi-\eta)^{1-\alpha-\beta}\!\!\!
\int_{\eta}^{\xi}\!\!\psi(a)(\xi-a)^{\beta-1}(a-\eta)^{\alpha-1}da
\nonumber\\
\!=\!{}&(\xi-\eta)^{1-\alpha-\beta}\!\!\!
\int_{0}^{1}\!\!\varphi(\xi+(\eta-\xi)t)t^{-\alpha}(1-t)^{-\beta}dt\nonumber\\
&\!+\!\!\int_{0}^{1}\!\!\psi(\xi+(\eta-\xi)t)t^{\beta-1}(1-t)^{\alpha-1}dt,
\label{eq:gensolED}
\end{align}
where $\varphi$ and $\psi$ are arbitrary functions and in the second equality
a new interpolation parameter has been introduced, $a=\xi+(\eta-\xi)t$. This
integral representation for the general solution was provided first for
$\alpha=\beta$ by Poisson \cite{Poisson:1823}, and then generalized to
$\alpha\neq\beta$ by Appell \cite{Appell:1882}.

The case $\alpha+\beta=1$ is not covered by the expression \eqref{eq:gensolED};
we cannot use the relation \eqref{eq:firstrel} to construct a second independent
particular solution. However, interestingly enough, we can use the final form
\eqref{eq:gensolED} to derive the general solution of this case as a
nontrivial limit. First, we notice that the generic solution
\eqref{eq:gensolED}  can be rewritten in the following form
\begin{align}
\mathfrak{u}_\text{g}={}&(\xi-\eta)^{1-\alpha-\beta}
\int_{0}^{1}dt\,t^{-\alpha}(1-t)^{-\beta}
\Biggl(\tilde{\varphi}(\xi+(\eta-\xi)t)\nonumber\\
&+\tilde{\psi}(\xi+(\eta-\xi)t)
\frac{ \left[(\xi-\eta)t(1-t)\right]^{\alpha+\beta-1}-1}
{\alpha+\beta-1}\Biggr),
\label{eq:quadraticRho}
\end{align}
where, without losing generality, we are considering that the arbitrary
functions could be defined from the beginning as
\[
\tilde{\varphi}=\varphi+\psi, \quad
\tilde{\psi}=(\alpha+\beta-1)\psi.
\]
The general solution to the Euler-Darboux equation when $\alpha+\beta=1$ is
precisely the limit of Eq.~\eqref{eq:quadraticRho} when the sum
$\alpha+\beta$ approaches unity; therefore
\begin{align}
\mathfrak{u}_\text{g}(\alpha,1-\alpha)={}&
\lim_{\beta \rightarrow 1-\alpha}\mathfrak{u}_\text{g}(\alpha,\beta)\nonumber\\
={}&\int_{0}^{1}dt\,t^{-\alpha}(1-t)^{\alpha-1}
\Bigl(\tilde{\varphi}(\xi+(\eta-\xi)t)\nonumber\\
&+\tilde{\psi}(\xi+(\eta-\xi)t)
\ln\left[(\xi-\eta)t(1-t)\right]\Bigr),
\label{eq:solgenED2}
\end{align}
which is just the general solution reported in Ref.~\cite{Koshlyakov} without
justification. This kind of solutions has turned out to be relevant in General
Relativity [specifically $\mathfrak{u}(1/2,1/2)$] in the construction of the
so-called Weyl class describing static axisymmetric vacuum spacetimes
\cite{Weyl:1917gp}, or in their analogues after a double Wick rotation characterizing
colliding plane gravitational waves with collinear polarization
\cite{Szekeres:1972uu}; see Refs. \cite{Stephani:2003,Griffiths:1991zp}
for more recent reviews on these subjects.

In what follows we emphasize there are extensions of the Euler-Darboux
equations \cite{Darboux:1899}
\begin{equation}\label{eq:extendedEDE}
\hat{E}_{\alpha,\beta;\lambda}(\mathfrak{U})\equiv
\left(E_{\alpha,\beta}+\frac{\lambda}{(\xi-\eta)^2}\right)\mathfrak{U}=0,
\end{equation}
which can be recast into the standard form. In the previous case, this is
done by just letting
\begin{equation}\label{eq:U2u}
\mathfrak{U}=(\xi-\eta)^{\rho}\mathfrak{u}.
\end{equation}
Indeed, $\mathfrak{u}$ must obey the standard Euler-Darboux equation
\begin{equation}
E_{\alpha+\rho,\beta+\rho}(\mathfrak{u})=0,
\end{equation}
provided that $\rho$ is a solution to the quadratic equation
\begin{equation}\label{eq:MassiveRelation}
\rho^2+(\alpha+\beta-1)\rho-\lambda=0.
\end{equation}
In this regard, it is important to make the following remark. In general, the
latter polynomial has two roots, say, $\rho_{\pm}$. Therefore, from
Eq. \eqref{eq:U2u} one may be tempted to write the general solution to the
extended Euler-Darboux equation \eqref{eq:extendedEDE} as a superposition
built from the contributions of both roots. However, it is enough to consider
a single root since, using relation \eqref{eq:firstrel}, it is easy to check
that
\begin{equation}
\mathfrak{u}(\alpha+\rho_{-},\beta+\rho_{-})
=(\xi-\eta)^{\sqrt{d}}\mathfrak{u}(\alpha+\rho_{+},\beta+\rho_{+}),
\end{equation}
where $d=(\alpha+\beta-1)^2-4\lambda$ is just the discriminant of the
quadratic equation \eqref{eq:MassiveRelation}. Hence, the general solution to
the extension \eqref{eq:extendedEDE} is the same when written with one root or the other
\begin{align}\label{eq:solmassEDE}
\mathfrak{U}
&=(\xi-\eta)^{\rho_{+}}\mathfrak{u}(\alpha+\rho_{+},\beta+\rho_{+})
\nonumber\\
&=(\xi-\eta)^{\rho_{-}}\mathfrak{u}(\alpha+\rho_{-},\beta+\rho_{-}).
\end{align}
For $\alpha=-1=\beta$, these are precisely the solutions of the massive
generalization of the Siklos equation we report in Sec.~\ref{Sec:GenSolSik}
and where the third parameter $\lambda$ is determined by the mass. More
relativistic examples for which the extended operator \eqref{eq:extendedEDE}
becomes relevant were reviewed in Ref.~\cite{Stewart:2009}, where almost all
aspects covered in this appendix and the following one were also reviewed from
a similar perspective.

Finally, we also need to address the problem of how to solve the
inhomogeneous Euler-Darboux equations
\begin{equation}\label{eq:InhEDe}
E_{\alpha,\beta}(\mathfrak{u})=f(\xi,\eta),
\end{equation}
since it is important in this paper to understand the behavior of AdS waves
with matter fields as sources. The inhomogeneous solutions of linear
hyperbolic equations such as these can be obtained by using the Riemann
method. In fact, Riemann devised his method by studying the
Euler-Poisson-Darboux equations \cite{Riemann:1860}. Hence, for
completeness{\color{blue},} we briefly review this method and how it must be
applied to the Euler-Darboux case in the last appendix.

\section{The Riemann Method \label{App:RG}}

Here we follow Ref.~\cite{Copson} but adopting a notation compatible with the
previous appendix. First of all, let us consider a linear second-order
differential operator
\begin{equation}\label{eq:LinearOpRiemann}
L(\mathfrak{u})=\left(
A\partial^2_{\xi\xi}+B\partial^2_{\eta\eta}+2C\partial^2_{\xi\eta}
+2D\partial_{\xi}+2E\partial_{\eta}+K\right)\mathfrak{u},
\end{equation}
with continuously differentiable coefficients. Multiplying the above by a
smooth function $\mathfrak{R}$ and differentiating by parts, the following
identity is obtained
\begin{subequations}
\begin{equation}\label{eq:DivergenceRG}
\mathfrak{R}L(\mathfrak{u})-\mathfrak{u}L^{*}(\mathfrak{R})=
\partial_\xi M+\partial_\eta N,
\end{equation}
where the functions determining the divergence on the right-hand side
\begin{align}
M={}&\partial_{\xi}(A\mathfrak{R}\mathfrak{u})
+\partial_{\eta}(C\mathfrak{R}\mathfrak{u})\nonumber\\
&-2\mathfrak{u}\bigl[\partial_\xi(A\mathfrak{R})
+\partial_\eta(C\mathfrak{R})
-D\mathfrak{R}\bigr],\label{eq:M=...}\\
N={}&\partial_\eta(B\mathfrak{R}\mathfrak{u})
+\partial_\xi(C\mathfrak{R}\mathfrak{u})\nonumber\\
&-2\mathfrak{u}\bigl[\partial_\eta(B\mathfrak{R})
+\partial_\xi(C\mathfrak{R})
-E\mathfrak{R}\bigr],\label{eq:N=...}
\end{align}
are not unique since we can add to them $\partial_\eta\Theta$ and
$-\partial_\xi\Theta$ respectively. However, the linear operator
\begin{align}\label{eq:AdjointOperator}
L^{*}(\mathfrak{R})={}&\partial^2_{\xi\xi}(A\mathfrak{R})
+\partial^2_{\eta\eta}(B\mathfrak{R})+2\partial^2_{\xi\eta}(C\mathfrak{R})
\nonumber\\
&-2\partial_{\xi}(D\mathfrak{R})
-2\partial_{\eta}(E\mathfrak{R})+K\mathfrak{R},
\end{align}
\end{subequations}
is unique and defines the \emph{adjoint} of $L$. Consequently, the adjoint of
$L^{*}$ is $L$. When additionally $L^{*}=L$ it is said that the operator $L$
is self-adjoint.

By integrating the identity \eqref{eq:DivergenceRG} in a well-defined domain
$\Omega$ whose boundary is a regular closed curve $\partial\Omega$ with
anticlockwise orientation, it follows from Green's theorem that
\begin{equation}\label{eq:GreenTheo}
\int\!\!\!\int_\Omega\left[\mathfrak{R}L(\mathfrak{u})
-\mathfrak{u}L^{*}(\mathfrak{R})\right]d\xi
d\eta=\int_{\partial\Omega}(Md\eta-Nd\xi).
\end{equation}
One of Riemann's significant contributions consists in an approach to solve
the Cauchy problem of an inhomogeneous hyperbolic equation using the above
Green integral identity \cite{Riemann:1860}. For a hyperbolic operator there
exist characteristic coordinates that allow to get rid of all of the second-order derivatives except the crossed one;
as a result, the associated inhomogeneous equation can be rewritten as
\begin{equation}\label{eq:HyperbolicEqn}
L(\mathfrak{u})=\left(\partial_{\xi\eta}+2D\partial_\xi
+2E\partial_\eta+K\right)\mathfrak{u}=f(\xi,\eta),
\end{equation}
where the inhomogeneity $f$ is a continuously differentiable functions, as are the coefficients $D$, $E$ and $K$.
Solving the Cauchy problem means finding the
unique solution to this equation with given values of $\mathfrak{u}$,
$\partial_\xi\mathfrak{u}$ and $\partial_\eta\mathfrak{u}$ on some initial
curve. Let $P=(\xi_0,\eta_0)$ be the point in the future (or past) of the
initial-value curve where it is required to know the solution. The first
step of the Riemann method is to choose the regular boundary $\partial\Omega$
of the identity \eqref{eq:GreenTheo} in such a way that it contains both the initial-value curve and the future (past)
point. In the characteristic plane
$(\xi,\eta)$ the initial-values curve must be a duly-inclined regular arc, and
we can define its intersections with the characteristics $\eta=\eta_0$ and
$\xi=\xi_0$ passing through $P$ as the points $Q=(\xi_1,\eta_0)$ and
$R=(\xi_0,\eta_1)$, respectively. Hence, the regular boundary
$\partial\Omega$ can be the deformed triangle with vertices $PQR$. The second
step of the method involves picking out the function $\mathfrak{R}$ in the identity \eqref{eq:GreenTheo} as a
homogeneous solution to the adjoint operator
\begin{subequations}\label{eq:RGFunction}
\begin{equation}\label{eq:*eq}
L^{*}(\mathfrak{R})=0,
\end{equation}
satisfying the additional conditions
\begin{align}
\partial_\xi\mathfrak{R}-2E\mathfrak{R}&=0
&\text{when}&&\eta&=\eta_{0},\label{eq:d_xiR}\\
\partial_\eta\mathfrak{R}-2D\mathfrak{R}&=0
&\text{when}&&\xi&=\xi_{0},\label{eq:d_etaR}\\
\mathfrak{R}(\xi_0,\eta_0;\xi_0,\eta_0)&=1.\label{eq:R=1}
\end{align}
\end{subequations}
This defines a characteristic boundary value problem whose solution
$\mathfrak{R}(\xi,\eta;\xi_0,\eta_0)$ is known as the \emph{Riemann-Green
function}.\footnote{This should not be confused with the well-known Green
function, which is an inhomogeneous solution for a Dirac delta source. For
the relationship between these functions, see Ref.~\cite{Mackie:1965}.} The
knowledge of this function for a hyperbolic linear partial differential
equation allows to construct the unique solution which is compatible with the
given initial conditions. Concretely, using the defining properties of
the Riemann-Green function \eqref{eq:RGFunction} in the identity \eqref{eq:GreenTheo}
after integrating at the deformed triangle with vertices $PQR$ and taking
into account the definitions \eqref{eq:M=...}--\eqref{eq:N=...}, it is possible to
isolate the value of the inhomogeneous solution $\mathfrak{u}$ at the point
$P=(\xi_0,\eta_0)$, i.e.\
\begin{subequations}\label{eq:uCauchy}
\begin{equation}\label{eq:u=uh+ui}
\mathfrak{u}(\xi_0,\eta_0)=\mathfrak{u}^\text{h}(\xi_0,\eta_0)
+\mathfrak{u}^\text{i}(\xi_0,\eta_0),
\end{equation}
where the first contribution solves the homogeneous equation respecting the
initial conditions and is given by
\begin{align}
\mathfrak{u}^\text{h}(\xi_0,\eta_0)={}&
\frac12\mathfrak{u}(\xi_0,\eta_1)
\mathfrak{R}(\xi_0,\eta_1;\xi_0,\eta_0)\nonumber\\
&+\frac12\mathfrak{u}(\xi_1,\eta_0)
\mathfrak{R}(\xi_1,\eta_0;\xi_0,\eta_0)\nonumber\\
&+\int_{QR}(Nd\xi-Md\eta),\label{eq:uh}
\end{align}
whereas the second contribution is a particular solution to the inhomogeneous
equation
\begin{equation}\label{eq:RGSol}
\mathfrak{u}^\text{i}(\xi_0,\eta_0)=
\int\!\!\!\int_{\Omega}f(\xi,\eta)\mathfrak{R}(\xi,\eta;\xi_{0},\eta_{0})
d\xi d\eta.
\end{equation}
\end{subequations}
An important remark is that in deriving the solution \eqref{eq:uCauchy} it has been
implicitly assumed that the initial value arc $QR$ is negatively inclined,
with the future point $P$ on its right, which is the more natural situation
for characteristic coordinates. However, using a different notation it is
also possible to have a positively inclined initial value arc with the future
point on its left. In this case, the deformed triangle $PQR$ is clockwise
oriented and as a consequence the solution is again expressed as
Eqs.~\eqref{eq:uCauchy}, except that the expression for the inhomogeneous solution
\eqref{eq:RGSol} changes sign. Returning to our main objective, if we know
the general homogeneous solution to Eq. \eqref{eq:HyperbolicEqn}, the first
contribution \eqref{eq:uh} is already considered there. Hence, we can rest in
the results of the Riemann approach and just keep the expression
\eqref{eq:RGSol}--or its negative counterpart in the positively inclined
case--to obtain a particular solution of the inhomogeneous hyperbolic
equation, as long as it is possible to build the corresponding Riemann-Green
function.

For our case of interest [the Euler-Darboux equation \eqref{eq:InhEDe}], the
first step is to find the corresponding adjoint operator that results in a
particular case of the extended Euler-Darboux operators
\eqref{eq:extendedEDE}, specifically
\begin{equation}\label{eq:E*=Extd}
E^*_{\alpha,\beta}(\mathfrak{R})=
\hat{E}_{-\beta,-\alpha;-\alpha-\beta}(\mathfrak{R})=0.
\end{equation}
Exploiting now the way in which the solutions of the extended and standard
Euler-Darboux equations are related [Eq.\eqref{eq:U2u}], the involved exponent is
determined from Eq.~\eqref{eq:MassiveRelation} and its two possible values are
$\rho=\{\alpha+\beta,1\}$. Since the result is the same regardless of which
one is elected, we choose $\rho=\alpha+\beta$ and the general homogeneous
solutions of the adjoint operator associated to the Euler-Darboux equations
can be written in terms of the solutions of the latter by just reversing their
parameters
\begin{equation}\label{eq:RED2u}
\mathfrak{R}=(\xi-\eta)^{\alpha+\beta}\mathfrak{u}(\beta,\alpha).
\end{equation}
It only remains to impose the boundary conditions
\eqref{eq:d_xiR}--\eqref{eq:R=1} after using, for example, the Poisson
integral representation \eqref{eq:gensolED} for the general solution of the
Euler-Darboux equations. The resulting Riemann-Green function acquires the
following form
\begin{subequations}\label{eq:RGfuncED}
\begin{align}
\mathfrak{R}(\xi,\eta;\xi_{0},\eta_{0})&=
\frac{(\xi-\eta)^{\alpha+\beta}}
{(\xi_{0}-\eta)^{\alpha}(\xi-\eta_{0})^{\beta}}\,
{_2F_1}(\alpha,\beta;1;\chi),\\
\chi&\equiv-\frac{(\xi-\xi_{0})(\eta-\eta_{0})}
{(\xi-\eta_{0})(\xi_{0}-\eta)},\label{eq:chi}
\end{align}
\end{subequations}
which can be derived in different ways \cite{Darboux:1899,Copson,Stewart:2009}, where ${_2F_1}$ stands for the standard
hypergeometric function.

Here, we provide an alternative derivation that allows us to prove at the same time
that Eq. \eqref{eq:RGSol} is in fact a particular solution of the inhomogeneous
Euler-Darboux equation \eqref{eq:InhEDe} if the above expression is
considered as the corresponding Riemann-Green function. Our main motivation
is to have a concrete expression available for the particular solution of the
complexified version we study in subsection~\ref{Subsubsec:Gms}, instead of
providing a general solution to the initial value problem. Hence, we consider
as a concrete initial value arc the straight line passing by the origin with
slope $-1$; this is equivalent to choosing $(\xi_1,\eta_1)=(-\eta_0,-\xi_0)$.
In order to make our derivation more transparent and also to be compatible
with the main text, we change the notation henceforth by relabeling the point
equipped with the Riemann solution using now standard coordinates, i.e.,\
$P=(\xi,\eta)$. Consequently, the other two vertices of the triangle forming
the integration region $\Omega$ are now given by $Q=(-\eta,\eta)$ and
$R=(\xi,-\xi)$. Additionally, we use primes to denote the dummy coordinates in
the integration. There are two elementary ways to sweep the triangle surface
$\Omega$ in the integration \eqref{eq:RGSol}: in the first, for each
horizontal coordinate of the triangle, we sum on the corresponding vertical
interval, and in the second we do exactly the opposite. We shall use a
superposition of both: they are equivalent in the real domain covered in this
appendix, but this is not the case if they are finally evaluated in complex
limits, as we pretend to do in the main text. The advantage of this
representation is that it naturally gives a real result in the complexified
context. Taking the above into account and inspired by Eqs.~\eqref{eq:RGSol} and
\eqref{eq:RGfuncED}, we propose as a particular solution to the inhomogeneous
Euler-Darboux equation \eqref{eq:InhEDe} the following expression
\begin{align}\label{eq:PSolED}
\mathfrak{u}^\text{i}(\xi,\eta)={}&\frac12\!\left(
\int_{-\eta}^\xi\!\! d\xi' \!\!\int_{-\xi'}^\eta\!\! d\eta'\!+
\!\int_{-\xi}^\eta\!\! d\eta'\!\!\int_{-\eta'}^\xi\!\! d\xi'\!\right)
\nonumber\\
&\times f(\xi',\eta')\frac{(\xi'-\eta')^{\alpha+\beta}}
{(\xi-\eta')^{\alpha}(\xi'-\eta)^{\beta}}F(\chi'),
\end{align}
where \emph{a priori} no other assumption is made on $F(\chi)$ except that it is a
function of the variable \eqref{eq:chi} written in the new notation. Applying
now the Euler-Darboux operator \eqref{eq:ede} to the above expression, we
obtain the following identity after a careful evaluation
\begin{subequations}\label{eq:IntIdentity}
\begin{align}
E_{\alpha,\beta}(\mathfrak{u}^\text{i})={}&F(0)f(\xi,\eta)\nonumber\\
&+\frac1{2(\xi-\eta)}\!\left(
\int_{-\eta}^\xi\!\! d\xi' \!\!\int_{-\xi'}^\eta\!\! d\eta'\!+
\!\int_{-\xi}^\eta\!\! d\eta'\!\!\int_{-\eta'}^\xi\!\! d\xi'\!\right)
\nonumber\\
&\times f(\xi',\eta')\frac{(\xi'-\eta')^{\alpha+\beta+1}}
{(\xi-\eta')^{\alpha+1}(\xi'-\eta)^{\beta+1}}\hat{H}_{\alpha,\beta;1}'(F),
\end{align}
where we denote the standard hypergeometric operator by
\begin{equation}
\hat{H}_{a,b;c}(F)\equiv\chi(\chi-1)F''+[(a+b+1)\chi-c]F'+abF.
\end{equation}
\end{subequations}
It is obvious now that if we choose the hitherto arbitrary function as the hypergeometric
\begin{equation}\label{eq:Hyperg}
F(\chi)={_2F_1}(\alpha,\beta;1;\chi),
\end{equation}
the integral contributions in the identity \eqref{eq:IntIdentity} vanish.
Considering now that ${_2F_1}(\alpha,\beta;1;0)=1$, it is proved that
Eq.~\eqref{eq:RGSol} [or, more specifically, Eq.~\eqref{eq:PSolED}] is in fact a
particular solution of the inhomogeneous Euler-Darboux equation
\eqref{eq:InhEDe} when the Riemann-Green function \eqref{eq:RGfuncED} is
considered. We emphasize that there are particular cases in the parameter
space for which it is even possible to integrate the expression \eqref{eq:PSolED} and arrive at a closed local form;
see subsection \ref{Subsubsec:Gms}.

Finally, for later use we write down the counterpart to Eq.~\eqref{eq:PSolED}
when a positive-inclined initial arc is considered. The concrete initial
value arc taken is the straight line passing by the origin with slope $1$,
which is equivalent to choosing $(\xi_1,\eta_1)=(\eta_0,\xi_0)$ in the notation
of the beginning of the appendix. In the notation of the last part, this
means integrating on the triangle with vertices $P=(\xi,\eta)$,
$Q=(\eta,\eta)$, and $R=(\xi,\xi)$, giving
\begin{align}\label{eq:PSolEDpi}
\mathfrak{u}^\text{i}(\xi,\eta)={}&-\frac12\!\left(
\int_{\xi}^\eta\!\! d\xi' \!\!\int_{\xi'}^\eta\!\! d\eta'\!+
\!\int_{\xi}^\eta\!\! d\eta'\!\!\int_{\xi}^{\eta'}\!\! d\xi'\!\right)
\nonumber\\
&\times f(\xi',\eta')\frac{(\xi'-\eta')^{\alpha+\beta}}
{(\xi-\eta')^{\alpha}(\xi'-\eta)^{\beta}}F(\chi').
\end{align}
The analog of the identity \eqref{eq:IntIdentity} also follows in this case, which
brings us to the same conclusion \eqref{eq:Hyperg}. With this, we end our
thorough review of results of the Euler-Darboux equations which are
indispensable to justify the more important findings of the main text.


\end{document}